\documentclass[12pt]{article}

\usepackage[auth-sc,affil-sl]{authblk}
\usepackage{graphicx}
\usepackage{color}
\usepackage{amsmath}
\usepackage{dsfont}
\usepackage[round]{natbib}
\usepackage{placeins}
\usepackage{amssymb}
\usepackage{abstract}
\usepackage{hyperref}
\usepackage{cancel}
\usepackage[margin=1.05in]{geometry}
\usepackage{enumerate}
\usepackage{listings}
\usepackage{mathdots}
\usepackage{array}
\usepackage{phaistos}
\usepackage{textcomp}
\usepackage{subcaption}
\usepackage{algorithm}
\usepackage{algorithmicx}
\usepackage{algpseudocode}

\newcommand{\qu}[1]{``#1''}

\newcommand{\treet}[1]{\text{\scriptsize \PHplaneTree}_{#1}}
\newcommand{\treeleaft}[1]{\text{\scriptsize \PHplaneTree}_{#1}^{\text{\tiny \textleaf}}}
\newcommand{\leaf}{\text{\scriptsize \textleaf}}

\newcounter{probnum}
\allowdisplaybreaks

\definecolor{gray}{rgb}{0.7,0.7,0.7}

\definecolor{black}{rgb}{0,0,0}
\definecolor{white}{rgb}{1,1,1}
\definecolor{blue}{rgb}{0,0,0.7}
\newcommand{\inblue}[1]{\color{blue}\textbf{#1} \color{black}}
\definecolor{green}{rgb}{0.133,0.545,0.133}
\newcommand{\ingreen}[1]{\color{green}\textbf{#1} \color{black}}
\definecolor{yellow}{rgb}{1,0.549,0}

\definecolor{red}{rgb}{1,0.133,0.133}
\newcommand{\inred}[1]{\color{red}\textbf{#1} \color{black}}
\definecolor{purple}{rgb}{0.58,0,0.827}

\definecolor{brown}{rgb}{0.55,0.27,0.07}

\definecolor{backgcode}{rgb}{0.97,0.97,0.8}
\definecolor{Brown}{cmyk}{0,0.81,1,0.60}
\definecolor{OliveGreen}{cmyk}{0.64,0,0.95,0.40}
\definecolor{CadetBlue}{cmyk}{0.62,0.57,0.23,0}

\newcommand{\bv}[1]{\boldsymbol{#1}}

\newcommand{\sigsq}{\sigma^2}

\newcommand{\bSigma}{\bv{\Sigma}}
\newcommand{\bSigmainv}{\bSigma^{-1}}

\newcommand{\musq}{\mu^2}

\newcommand{\bgamma}{\bv{\gamma}}

\newcommand{\transpose}[1]{\parens{#1}^{\top}}

\newcommand{\iid}{~{\buildrel iid \over \sim}~}
\newcommand{\inddist}{~{\buildrel ind \over \sim}~}

\newcommand{\half}{\frac{1}{2}}

\newcommand{\B}{\bv{B}}
\newcommand{\D}{\bv{D}}

\newcommand{\R}{\bv{R}}

\newcommand{\Z}{\bv{Z}}
\newcommand{\X}{\bv{X}}

\newcommand{\I}{\bv{I}}

\newcommand{\0}{\bv{0}}

\newcommand{\Y}{\bv{Y}}

\newcommand{\x}{\bv{x}}
\newcommand{\onevec}{\bv{1}}

\newcommand{\y}{\bv{y}}

\newcommand{\z}{\bv{z}}

\newcommand{\threebythreemat}[9]{\bracks{\begin{array}{ccc} #1 & #2 & #3 \\ #4 & #5 & #6 \\ #7 & #8 & #9 \end{array}}}

\newcommand{\reals}{\mathbb{R}}

\newcommand{\beqn}{\vspace{-0.25cm}\begin{eqnarray*}}
\newcommand{\eeqn}{\end{eqnarray*}}
\newcommand{\bneqn}{\vspace{-0.25cm}\begin{eqnarray}}
\newcommand{\eneqn}{\end{eqnarray}}

\newcommand{\parens}[1]{\left(#1\right)}
\newcommand{\squared}[1]{\parens{#1}^2}
\newcommand{\tothepow}[2]{\parens{#1}^{#2}}
\newcommand{\prob}[1]{\mathbb{P}\parens{#1}}
\newcommand{\cprob}[2]{\prob{#1~|~#2}}

\newcommand{\sumonen}[2]{\sum_{#1=1}^n #2}

\newcommand{\bracks}[1]{\left[#1\right]}
\newcommand{\braces}[1]{\left\{#1\right\}}

\newcommand{\abss}[1]{\left|#1\right|}

\newcommand{\inverse}[1]{\parens{#1}^{-1}}

\renewcommand{\exp}[1]{\mathrm{exp}\parens{#1}}

\newcommand{\natlog}[1]{\ln\parens{#1}}
\newcommand{\oneover}[1]{\frac{1}{#1}}
\newcommand{\overtwo}[1]{\frac{#1}{2}}

\newcommand{\oneoversqrt}[1]{\oneover{\sqrt{#1}}}

\newcommand{\multnormnot}[3]{\mathcal{N}_{#1}\parens{#2,\,#3}}
\newcommand{\normnot}[2]{\mathcal{N}\parens{#1,\,#2}}

\newcommand{\uniform}[2]{\mathrm{U}\parens{#1,\,#2}}

\newcommand{\invgammanot}[2]{\text{InvGamma}\parens{#1,\,#2}}

\newcommand{\normpdf}[3]{\frac{1}{\sqrt{2\pi#3}}\exp{-\frac{1}{2#3}(#1 - #2)^2}}

\newcommand{\normpdfmeanzero}[2]{\frac{1}{\sqrt{2\pi#2}}\exp{-\frac{1}{2#2}#1^2}}

\newcommand{\zeroonecl}{\bracks{0,1}}

\newcommand{\myint}[4]{\int_{#2}^{#3} #4 \,\text{d}#1}

\newcommand{\errorrv}{\mathcal{E}}
\newcommand{\berrorrv}{\bv{\errorrv}}

\newcommand{\transp}[1]{\parens{#1}^\top}

\newcommand{\doneover}[1]{\dfrac{1}{#1}}

\newcommand{\sigsqmu}{\sigsq_\mu}

\title{Bayesian Additive Regression Trees With Parametric Models of Heteroskedasticity}
\author{Justin Bleich\footnote{Both authors contributed equally to this work.}~\thanks{Electronic address: \texttt{jbleich@wharton.upenn.edu}; Prinicipal Corresponding author}~}
\author{Adam Kapelner$^*$\thanks{Electronic address: \texttt{kapelner@wharton.upenn.edu}; Corresponding author}}
\affil{The Wharton School of the University of Pennsylvania}

\begin{document}
\maketitle

\begin{abstract}
We incorporate heteroskedasticity into Bayesian Additive Regression Trees (\texttt{BART}) by modeling the log of the error variance parameter as a linear function of prespecified covariates. Under this scheme, the Gibbs sampling procedure for the original sum-of-trees model is easily modified, and the parameters for the variance model are updated via a Metropolis-Hastings step. We demonstrate the promise of our approach by providing more appropriate posterior predictive intervals than homoskedastic \texttt{BART} in heteroskedastic settings and demonstrating the model's resistance to overfitting. Our implementation will be offered in an upcoming release of the \texttt{R} package \texttt{bartMachine}.
\end{abstract}

\section{Introduction}

We consider the the following general heteroskedastic regression framework to characterize the relationship between a continuous response vector $\y$ and a set of $p$ predictor variables  $\X := \bracks{\x_{\cdot 1},\ldots,\x_{\cdot p}}$ which can be continuous or categorical:

\beqn
\y = f(\X) + \berrorrv, \quad\quad \berrorrv \sim \multnormnot{n}{\0}{\sigsq\D} 
\eeqn

\noindent $\D$ denotes the diagonal matrix whose entries are scaling factors for the error variance for each observation. In this model, the response is considered an unknown function $f$ of the predictors and the observations, while independent, exhibit non-constant error variance. The goal of this article is to model the relationship between the predictors and response with the aim of generating accurate predictions. To this end we model $f$ with Bayesian Additive Regression Trees \citep[\texttt{BART},][]{Chipman2010} which is a sum-of-trees model that has demonstrated predictive performance competitive with the best statistical learning algorithms. The original \texttt{BART} model is constrained to homoskedastic error variance ($\D = \I$). Here, we extend the model to flexibly handle an error variance structure which is a linear model of prespecified covariates and we name our procedure \qu{heteroskedastic \texttt{BART}} or \qu{\texttt{HBART}.} Similar to Huber-White sandwich estimation \citep{White1980}, appropriately modeling the diagonal entries of $\D$ \qu{downweights} high variance observations. This allows for (a) a more accurate model as measured by predictive performance on future observations as well as (b) posterior credible and predictive intervals which appropriately reflect the changing heteroskedasticity in predictor space.

In Section \ref{sec:lit_review}, we provide an overview of the literature on heteroskedastic regression modeling in a Bayesian paradigm. In Section~\ref{sec:HBART}, we introduce \texttt{HBART}, highlighting the necessary modifications to the original homoskedastic \texttt{BART}. In Section~\ref{sec:sims}, we provide simulations to showcase the desirable properties of \texttt{HBART}, including less overfitting for high noise observations as well as more appropriate uncertainty intervals for predictions in the presence of heteroskedasticity. Section~\ref{sec:real_data} explores two applications to real data. We conclude and offer future research directions in Section~\ref{sec:discussion}. The method developed in this paper will be implemented in an upcoming release of the \texttt{R} package \texttt{bartMachine} \citep{Kapelner2013}, which is available on CRAN.

\section{Bayesian Heteroskedastic Regression}\label{sec:lit_review}

Early approaches for heteroskedastic regression primarily focused on point estimation for the parameters governing the underlying heteroskedasticity of the model (for an overview, see \citealp{Carroll1988}). Potential problems with point estimation gave rise to a proposal of a fully Bayesian approach for heteroskedastic linear regression, where the non-constant variance depends on simple functions of an unknown parameter $\theta$ and a set of weights $w_i$ \citep{Boscardin1994}. 

\citet{Cepeda2001} introduce a Bayesian regression model where the conditional mean of the response is modeled as a linear function of covariates $\x_1, \ldots, \x_p$ plus heteroskedastic noise. They model the variance for each observation as a monotonic differentiable function of a linear combination of another set of covariates, $g(\z_1, \ldots, \z_{k})$. Additionally, the function $g$ is chosen to ensure positivity of the variance terms. The authors rely on a block Gibbs sampling approach \citep{Geman1984}, sampling the parameters for the mean function and variance function in two stages. In particular, the parameters for the variance function are updated via a Metropolis-Hastings step \citep{Hastings1970} using the approach of \citet{Gamerman1997}, which relies on an iteratively reweighted least squares model to generate suitable proposal distributions. 

More recent approaches have focused on relaxing the assumptions of linear additive components for modeling the mean and variance functions. \citet{Yau2003} propose nonparametric models for each of these two functions by employing penalized regression spline estimation for both models. \citet{Chan2006} extend this nonparametric model to allow for semiparametric modeling of both the mean and variance functions, using radial basis functions for nonparametric components. Additionally, their approach can handle a large number of basis terms by introducing Bayesian variable selection priors, allowing model estimation to be locally adaptive. \citet{Leslie2007} relax the assumption of normal errors and developed a heteroskedastic linear regression model with general error distributions by relying on a Dirichlet process mixture prior. 

Both \citet{Chan2006} and \citet{Leslie2007} rely on the sampling scheme developed in \citet{Gamerman1997} to obtain draws from the posterior distribution of the parameters for the variance function. Our work similarly draws heavily on this technique.

\section{Augmenting \texttt{BART} to Incorporate Heteroskedasticity}\label{sec:HBART}

In its original formulation in \citet{Chipman2010}, the authors assume that the response $\y$ could be modeled as a sum-of-trees model of the covariates $\X := \bracks{\x_{\cdot 1},\ldots,\x_{\cdot p}}$ plus homoskedastic normal noise:

\bneqn\label{eq:homo_bart}
\Y = \sum\limits_{i = 1}^{m} \treeleaft{i} (\X) + \berrorrv, \quad\quad \berrorrv \sim \multnormnot{n}{\0}{\sigsq \bv I_n} \eneqn

Each of the $m$ distinct binary regression trees consists of a tree structure, denoted by $\treet{}$, and a set of parameters in the terminal nodes of the tree (also called \qu{leaves}), denoted by $\leaf$. A tree with both its structure and set of leaf parameters is denoted by $\treeleaft{}$.

The structure of a given tree $\treet{t}$ includes information on how any observation recurses down the tree until a leaf node is reached. Each nonterminal, internal node contains a \qu{splitting rule} $\x_j < c$. When this condition is satisfied for a given observation, the observation moves to the left daughter node or moves to the right daughter node otherwise. Once an observation lands in a terminal node, the leaf value in that node is assigned as the predicted value for that observation. $\leaf_t = \braces{\mu_{t,1}, \mu_{t,2}, \ldots, \mu_{t_{b_t}}}$ represents the set of leaf parameters for a given tree, where $b_t$ is the number of terminal nodes in that tree.

Thus, \texttt{BART} estimates the mean function $f$ using a sum-of-regression trees. Regression trees rely on recursive binary partitioning of predictor space into a menagerie of hyperrectangles in order to approximate the unknown function of interest. By employing a sum-of-trees approach, \texttt{BART} is able to take advantage of the regression trees' ability to successfully model nonlinearities and interactions while also capturing additive components of the fit. As a Bayesian model, \texttt{BART} is composed of a set of priors and a likelihood. Using Gibbs sampling, posterior inference for the unknown $f$ can be obtained.

We propose an extension to \texttt{BART} by allowing each $\sigsq_1, \ldots, \sigsq_n$ to be scaled by the exponential of a linear parametric function of $k$ covariates $\Z := \bracks{\z_{\cdot 1}, \ldots, \z_{\cdot k}}$, the \qu{heteroskedasticity covariates} which are potentially distinct from the covariates used to model the mean function, $\x_{\cdot 1}, \ldots, \x_{\cdot p}$. Our heteroskedastic model, \texttt{HBART}, is given as

\bneqn\label{eq:hetero_bart_model}
\y & = & \sum\limits_{i=1}^{m} \treeleaft{i}(\X) + \berrorrv, \quad \berrorrv \sim \multnormnot{n}{\0}{\sigsq\threebythreemat{\exp{\z_{1 \cdot} \cdot \bgamma}}{}{0}{}{\ddots}{}{0}{}{\exp{\z_{n \cdot} \cdot \bgamma}}}
\eneqn

\noindent where $\bgamma := \bracks{\gamma_1, \ldots, \gamma_k}^\top$ is a column vector of linear coefficients for the $k$ heteroskedasticity covariates. Thus, the variance of each observation is specified as a log-linear model: 

\bneqn\label{eq:log_linear_model}
\natlog{\sigsq_i} = \natlog{\sigsq} + \z_i \cdot \bgamma \quad \text{for} \quad i = 1,\ldots,n.
\eneqn

It is important to note that \texttt{BART}, by design, is an overparameterized model with \qu{an abundance of unidentified parameters} \citep{Chipman2010} allowing for a highly flexible fit. First, given the unidentifiable nature of the model, our focus is not on valid inference for $\bgamma$. Instead, we incorporate heteroskedasticity to aid in forecasting and generating posterior uncertainty intervals. Second, due to the already complex nature of the original \texttt{BART} algorithm, we employ parametric models for heteroskedasticity versus more sophisticated alternatives (such as the proposal of \citealp{Chan2006}) in order to prevent the model from becoming \qu{too flexible.} Given highly flexible, unidentifiable estimation of both the mean and variance functions, the algorithm may have difficulty distinguishing between signal and noise, thereby shirking on its primary duty which is accurate estimation of the mean model. Hence, parametric models of heteroskedasticity represent the first step towards understanding how flexible \texttt{BART} can be in nonparametric function estimation when the homoskedasticity assumption is relaxed.

The remainder of the section is dedicated to describing the priors on \texttt{HBART} as well as the Gibbs sampling procedure for obtaining posterior inference. 

\subsection{Priors and Likelihood}\label{subec:priors}

The \texttt{BART} model requires three priors. The first prior is on the tree structures themselves and the second is on the leaf parameters. The third prior is on the error variance $\sigsq$. \texttt{HBART} requires these same three priors as well as a prior on $\bv \gamma$. By assumption, the priors on $\sigsq$, $\gamma$, $\treet{}$ and $\leaf$ are independent of one another:

\beqn 
\prob{\treeleaft{1},\ldots,\treeleaft{m}, \sigsq, \bv \gamma} &=& \bracks{\prod_{t}\prob{\treeleaft{t}}}\prob{\sigsq}\prob{\bv \gamma} 
\\ &=& \bracks{\prod_{t}\cprob{\leaf_t}{\treet{t}}\prob{\treet{t}}}\prob{\sigsq}\prob{\bv \gamma} \\
&=& \bracks{\prod_{t}\prod_{\ell}\cprob{ \mu_{t,\ell}}{\treet{t}}\prob{\treet{t}}}\prob{\sigsq}\prob{\bv \gamma}
\eeqn

\noindent where the last line follows from an assumption of conditional independence of the leaf parameters given the tree structure.

The prior on the tree structures and leaf parameters are designed to provide regularization to the model by respectively preventing the trees from growing too deep and shrinking the predicted values towards the range center of the response. These priors, as well as a prior on proposing splitting rules, are the same as those used in the original \texttt{BART} model and we refer the interested reader to \citet{Chipman2010}. 

For the homoskedastic implementation, the prior is on the error variance and is chosen to be $\sigsq \sim \invgammanot{\nu / 2}{\nu\lambda / 2}$. $\lambda$ is determined from the data so that there is a $q = 90\%$ a priori chance (by default) that the \texttt{BART} model will improve upon the root mean square error (RMSE) from an ordinary least squares regression (therefore, the majority of the prior probability mass lies below the RMSE of a least squares regression). Additionally, this prior limits the probability mass placed on small values of $\sigsq$ to prevent overfitting. We use this same data-informed prior for the $\sigsq$ parameter in \texttt{HBART}, the logarithm of which serves as the intercept term in the log-linear model for the variances (Equation~\ref{eq:log_linear_model}).

Following the approach of \citet{Gamerman1997}, we place a multivariate normal prior on $\bv \gamma$ given as:

\bneqn\label{eq:gamma_prior}
\bv \gamma &\sim& \multnormnot{k}{\bv\gamma_0}{\threebythreemat{\Sigma_{11}}{}{0}{}{\ddots}{}{0}{}{\Sigma_{kk}}}
\eneqn

We make the simplifying assumption that each component of the prior is independent of one another, but this can be easily generalized.

Along with a set of priors, \texttt{HBART} (and \texttt{BART}) consists of the likelihood of responses in the leaf nodes. The likelihood is assumed to be normal with the mean being the \qu{best guess} of the leaf parameters at the current moment and variance being the best guess of the variance at the moment i.e. $\y_\ell \sim \normnot{\mu_\ell}{\sigsq_i / m}$. These \qu{best guesses} are the values being conditioned on in the Gibbs sampler during each iteration. 

\subsection{Sampling from the Posterior}\label{subsec:posterior_sampling}

The Gibbs sampler can be used to obtain draws from $\mathbb{P}(\treeleaft{1}, \ldots, \treeleaft{m}, \sigsq, \bv\gamma ~|~\y, \X)$, the posterior distribution of the model parameters. As with the original sampling scheme for \texttt{BART}, \texttt{HBART} also relies on \qu{Bayesian backfitting} \citep{Hastie2000} to fit each tree iteratively, holding all other $m - 1$ trees constant. This is achieved by considering the residual response when updating the $j^{th}$ tree $\R_j := \y - \sum_{t \neq j} \treeleaft{t}(\X)$.

The Gibbs sampler for \texttt{HBART} proceeds by first proposing a change to the first tree's structure $\treet{1}$ which are accepted or rejected via a Metropolis-Hastings step \citep{Hastings1970}. Tree structures are altered by introducing small changes: growing a terminal node by adding two terminal daughter nodes, pruning two terminal daughter nodes, or changing a split rule. We denote these possible alterations as: GROW, PRUNE, and CHANGE.\footnote{An additional step known as SWAP was proposed in \citet{Chipman2010}, but this step is not implemented in the \texttt{bartMachine} package which was used to develop \texttt{HBART}.} Given the tree structure, samples from the posterior of the $b$ leaf parameters $\leaf_1 := \braces{\mu_1, \ldots, \mu_b}$ are then drawn from the conjugate-normal posterior distribution. This procedure proceeds iteratively for each tree, using an updated set of partial residuals $\R_j$. 

Once each tree structure and leaf values has been updated, a draw from the posterior of $\sigsq$ conditional on all other parameters is made based on the full model residuals $\berrorrv := \y - \sum_{t = 1}^m \treeleaft{t}(\X)$. Finally, a draw from the posterior of $\bgamma$ conditional the other parameters is computed via a Metropolis-Hastings step. 

The steps of the full procedure are illustrated below:

\bneqn \label{eq:gibbs_sampler}
1: & \treet{1} &|~~ R_{-1}, \sigsq, \bgamma \\
2:& \leaf_1 &|~~ \treet{1}, R_{-1}, \sigsq, \bgamma \nonumber\\ 
&\vdots \nonumber\\
2m - 1:& \treet{m} &|~~ R_{-m}, \sigsq, \bgamma\nonumber\\
2m:& \leaf_m &|~~ \treet{m}, R_{-m}, \sigsq, \bgamma \nonumber\\ 
2m + 1:& \sigsq &|~~ \treeleaft{1}, \ldots, \treeleaft{m}, \bgamma, \berrorrv \nonumber\\
2m + 2:& \bgamma &|~~ \sigsq, \berrorrv \nonumber
\eneqn

All $2m+2$ steps represent a \textit{single} Gibbs iteration\footnote{\texttt{BART} relies on a similar scheme, removing the conditioning on $\bv \gamma$ at each step and not requiring step $2m +2$.}. After a sufficient burn-in period, burned-in draws from the posterior of $f$ are obtained. A point prediction $\hat{f}(\x)$ can be obtained by taking the average of the burned-in values of $\treeleaft{1}, \ldots, \treeleaft{m}$ evaluated at a given $\x$; posterior credible intervals for $f$ are computed by using the quantiles of the burned-in values. Posterior predictive intervals at a given $\x$ are computed as follows:

\begin{enumerate}[1.]
\item Draw $f(\x)$ via
\begin{enumerate}[i)]
\item drawing one of the burned-in sum-of-trees collections $\treeleaft{1}, \ldots, \treeleaft{m}$ and
\item computing $f(\x) = \sum\limits_{i = 1}^{m} \treeleaft{i} (\x)$.
\end{enumerate}
\item Draw $\sigsq(\x)$ via
\begin{enumerate}[i)]
\item obtaining  $\bgamma$ and $\sigsq$ from the same Gibbs sample from which the sum-of-trees was obtained,
\item determining the $\z$ which corresponds to the $\x$ of interest and
\item computing $\sigsq(\x)$ by evaluating $\bgamma,~\sigsq$ and $\z$ in the exponentiation of Equation~\ref{eq:log_linear_model}.
\end{enumerate}

\item Take one draw from $\mathcal{N}({f}(\x), \sigsq(\x))$ which is the BART model (Equation~\ref{eq:homo_bart})
\item Collect draws from step 3 by repeating steps 1--3 many times. Then, return the desired quantiles.
\end{enumerate}

Note that \texttt{HBART} requires modifications to the original \texttt{BART} likelihood calculations necessary for Metropolis-Hastings steps to alter the tree structures (steps $1, 3, \ldots, 2m - 1$ of Equation~\ref{eq:gibbs_sampler}). Also, the posterior distributions for the leaf parameters must be updated (steps $2, 4, \ldots, 2m$ of Equation~\ref{eq:gibbs_sampler}). It is worth noting that these modifications are valid for any heteroskedastic \texttt{BART} model and not just the log-linear \texttt{HBART} model proposed in this work, as they computed as functions of $\sigsq_1, \ldots, \sigsq_n$. Additionally, modifications to the posterior distribution of $\sigsq$ (step $2m + 1$ of Equation~\ref{eq:gibbs_sampler}) is required. Finally, the sample of $\bgamma$ (step $2m + 2$ of Equation~\ref{eq:gibbs_sampler}) is obtained using the Metropolis-Hastings procedure with the proposal distribution outlined in \citet{Gamerman1997}. We provide explicit computational details for each of these steps in Appendix~\ref{app:details}.

\section{Simulations}\label{sec:sims}

\subsection{Univariate Model}\label{subsec:univariate_sim}

We begin by comparing the performance of \texttt{BART} versus \texttt{HBART} in a simple univariate setting. Consider a single predictor $\x$ which is a uniformly spaced sequence of $n$ points on $\zeroonecl$. Then, we consider two models. The first is heteroskedastic and is given by:

\bneqn\label{eq:simple_hetero}
Y_i = 100x_i + \errorrv_i, \quad \errorrv_i \inddist \normnot{0}{\sigsq_i}, \quad \sigsq_i = \exp{7x_i}
\eneqn

\noindent The second is homoskedastic and is given by:

\bneqn\label{eq:simple_homo}
Y_i = 100x_i + \errorrv_i, \quad \errorrv_i \inddist \normnot{0}{5^2}
\eneqn

For \texttt{HBART}, the matrix of covariates for the parametric variance model will be set to $\Z = \bracks{\x}$, our one uniformly spaced covariate. Parenthetically, note that the default in our software implementation is to set $\Z = \X$. Therefore, \texttt{HBART} will have a correctly specified variance function for the model given in Equation~\ref{eq:simple_hetero}. We include the homoskedastic model as well to benchmark \texttt{HBART}'s performance to determine if the unnecessary extra complexity of the variance model degrades the algorithm's performance. 

\begin{figure}[htp]
\centering
\begin{subfigure}[b]{0.49\textwidth}
\centering
\includegraphics[width=3in]{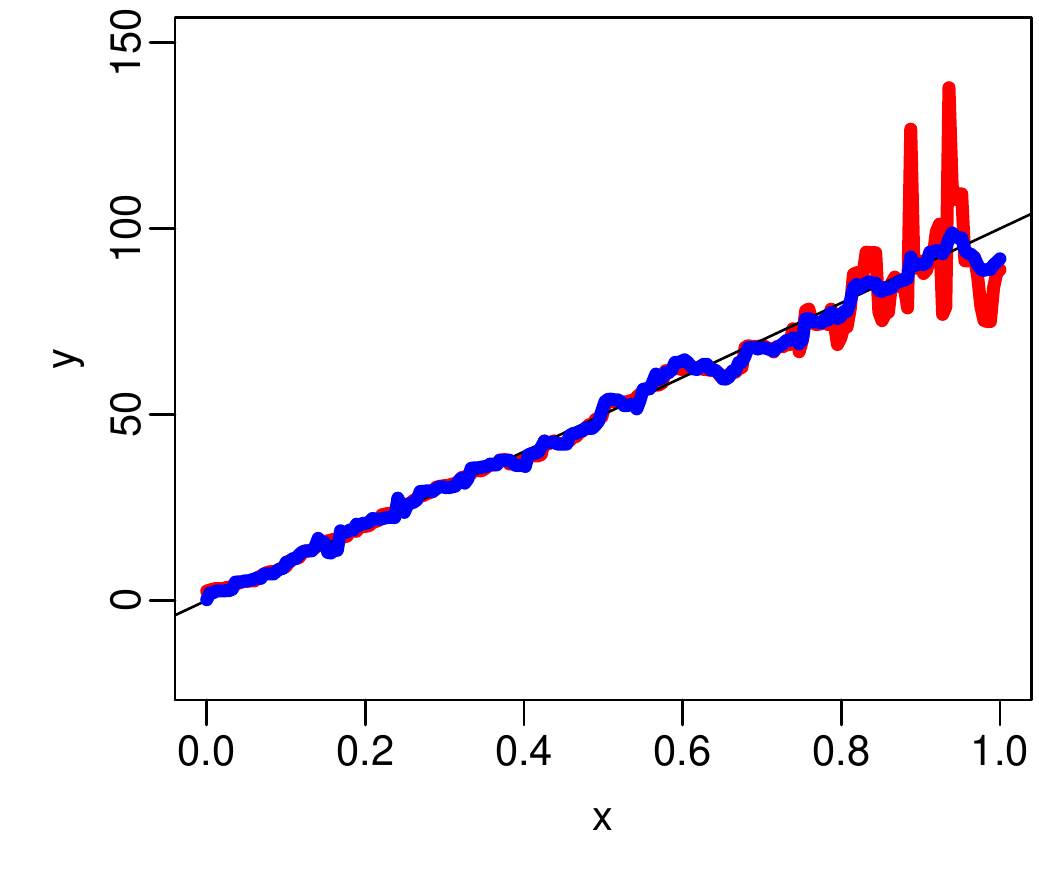}
\caption{\centering Heteroskedastic Model, \linebreak Posterior Mean Estimates}
\label{subfig:mean_estimation}
\end{subfigure}~
\begin{subfigure}[b]{0.49\textwidth}
\centering
\includegraphics[width=3in]{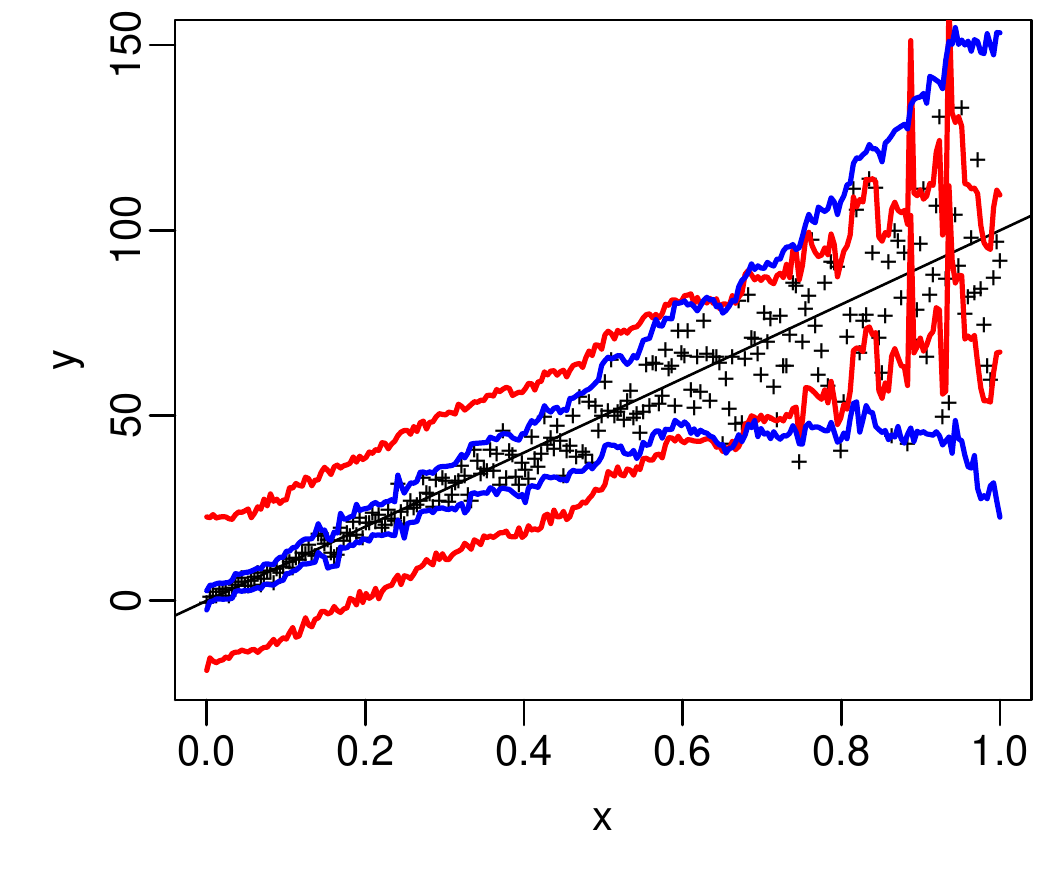}
\caption{\centering Heteroskedastic Model, \linebreak Poseterior Predictive Intervals}
\label{subfig:pred_intervals}
\end{subfigure}\\
\begin{subfigure}[b]{0.49\textwidth}
\centering
\includegraphics[width=3in]{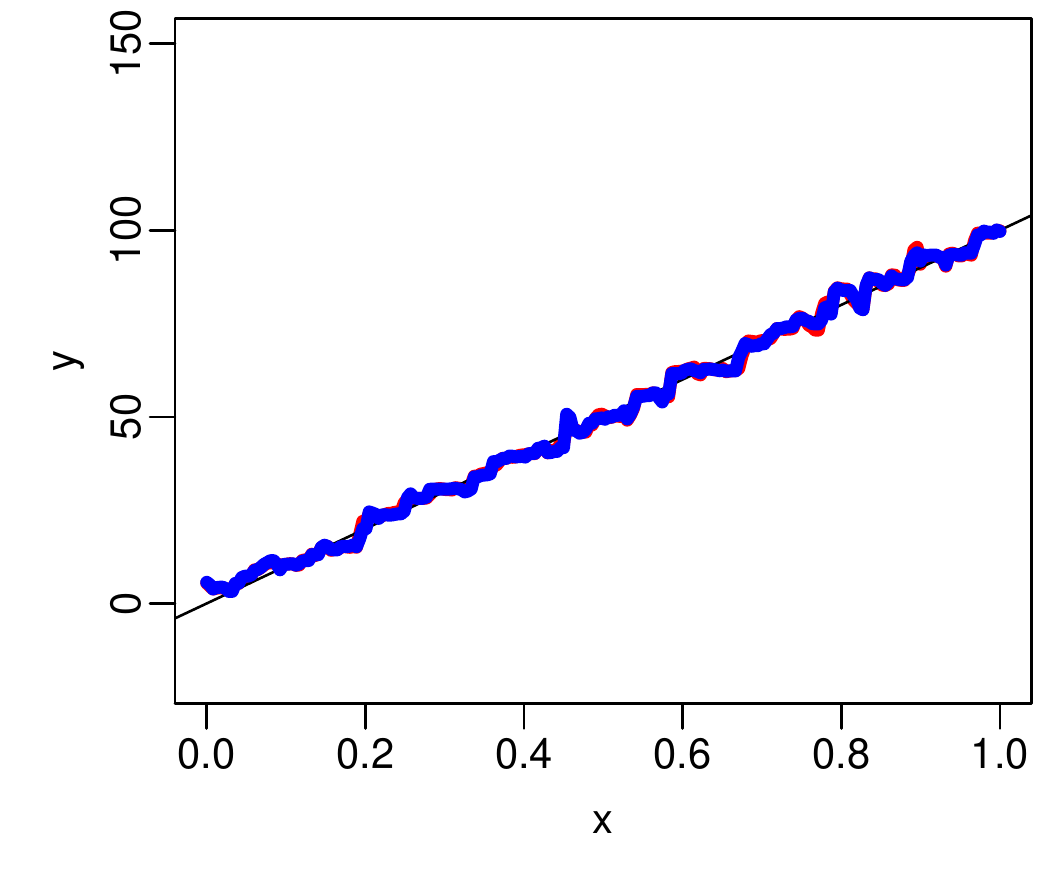}
\caption{\centering Homoskedastic Model, \linebreak Posterior Mean Estimates}
\label{subfig:mean_estimation_homo}
\end{subfigure}~
\begin{subfigure}[b]{0.49\textwidth}
\centering
\includegraphics[width=3in]{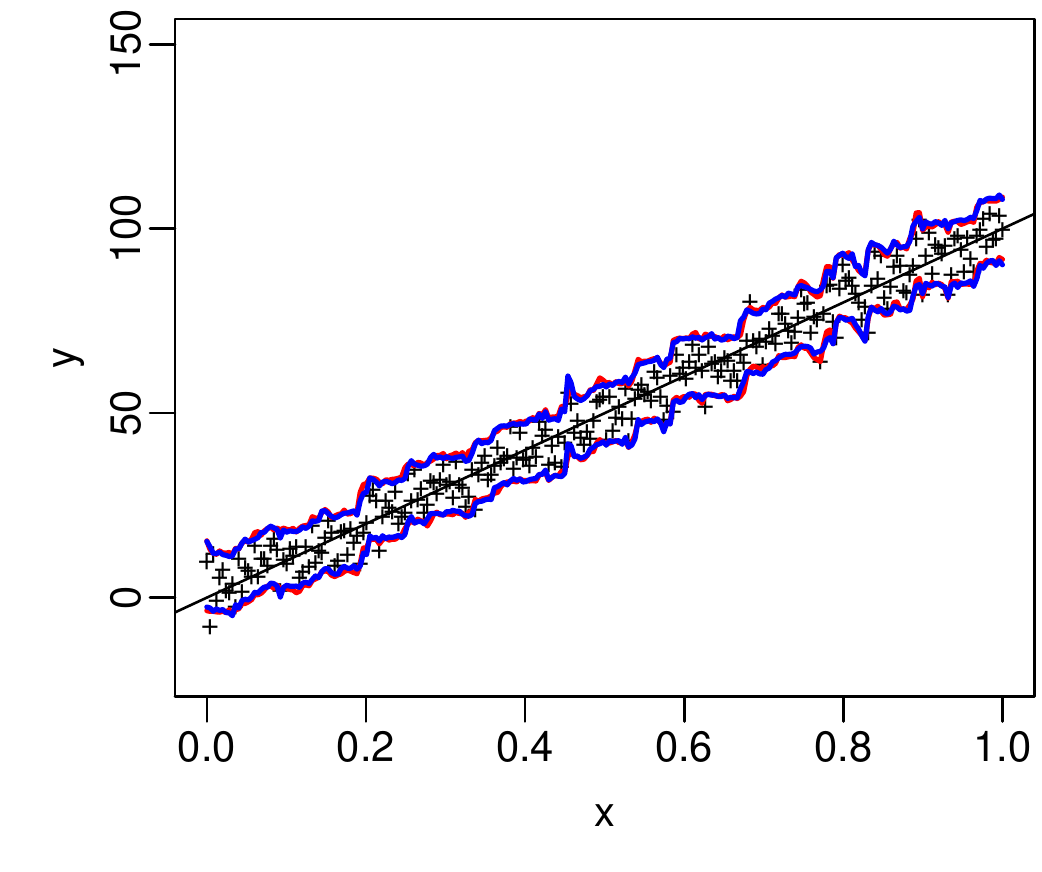}
\caption{\centering Homoskedastic Model, \linebreak Posterior Predictive Intervals}
\label{subfig:pred_intervals_homo}
\end{subfigure}
\caption{\texttt{BART}'s and \texttt{HBART}'s posterior mean estimates and 90\% posterior predictive intervals for each algorithm built from a sample of $n=250$ observations drawn from the processes in Equation~\ref{eq:simple_hetero} (a and b) and Equation~\ref{eq:simple_homo} (c and d). \inred{Red} lines correspond to the results of a \texttt{BART} model and \inblue{blue} lines correspond to the results from a \texttt{HBART} model. The black lines in (b) and (d) correspond to the true conditional mean function ($Y_i = 100 x_i$) and the black $+$'s represent the actual observations.}
\label{fig:univariate_sim}
\end{figure}

Figure~\ref{fig:univariate_sim} compares \texttt{BART} to \texttt{HBART} in both the heteroskedastic and homoskedastic simulated models by gauging two metrics: accuracy of estimation of $f$ and appropriateness of posterior predictive intervals.

Figure~\ref{subfig:mean_estimation} highlights the posterior means $\hat{f}$ estimated by the two different models. Note that \texttt{HBART} provides a better estimate of the true $f$ when $x > 0.7$, the region of relatively high variance. By estimating large variance in this region, it has the flexibility to downweight these high variance observations, allowing for more shrinkage towards the global average and away from the noisy local (within-node) sample mean. In contrast, \texttt{BART} overfits in this region. By assuming homoskedasticity, the algorithm is handicapped, and is obligated to move its mean function up when the noise term is large and positive and down when the noise term is large and negative. Note that both algorithms perform well when $x < 0.7$, where the data has lower variance.

Figure~\ref{subfig:pred_intervals} provides 90\% posterior predictive intervals for future observations (as explained in the procedure outlined in Section~\ref{subsec:posterior_sampling}). Given the homoskedasticity assumption of \texttt{BART}, the prediction intervals at each $x$-location are of constant width. This implies that the intervals are too wide at low values of $x$ and too narrow at higher values of $x$. In contrast, \texttt{HBART} provides more appropriate prediction intervals, narrow at low $x$ and wide at high $x$, thus correctly reflecting the heteroskedasticity in the underlying data-generating model. Although not the primary focus of this paper, examining the burned in $\gamma$ values yielded a 90\% credible interval of $[6.52, 7.65]$, which captures 7, the value of the linear coefficient in the log-linear variance model of Equation~\ref{eq:simple_hetero}.

For the homoskedastic model, Figures~\ref{subfig:mean_estimation_homo} and \ref{subfig:pred_intervals_homo} highlight that \texttt{BART} and \texttt{HBART} yield virtually identical results in terms of mean function estimation and predictive intervals. Thus, \texttt{HBART} seems to be robust in the absence of heteroskedasticity for this illustration.

\subsection{Multivariate Model}\label{subsec:multivariate_sim}

We now consider the following data generating process, similar to the model simulated in \citet{Cepeda2001}: 

\bneqn
&& Y_i =f(\x_i) + \errorrv_i, \quad f(\x_i) = -35 + .35x_{1,i} -1.7 x_{2,i}, \quad \errorrv_i \iid \normnot{0}{\sigsq_i} \label{eq:multi_model}\\
&& x_{1,i}\iid \uniform{0}{400}, \quad x_{2,i}\iid \uniform{10}{23}, \quad x_{3,i}\iid \uniform{0}{10} \nonumber \\
&& \sigsq_i = \exp{-6 + .03x_{1,i} + .4x_{3,i}} \label{eq:multi_model_var}
\eneqn

We set the number of observations to be $n=500$ and then we fit an \texttt{HBART} model using the default $\Z = \bracks{\x_1, \x_2, \x_3}$. Thus, the variance model is misspecified (the true model is not a function of $\x_2$). We again compare \texttt{HBART}'s performance to that of \texttt{BART}.

\begin{figure}[htp]
\centering
\begin{subfigure}[b]{0.49\textwidth}
\centering
\includegraphics[width=3in]{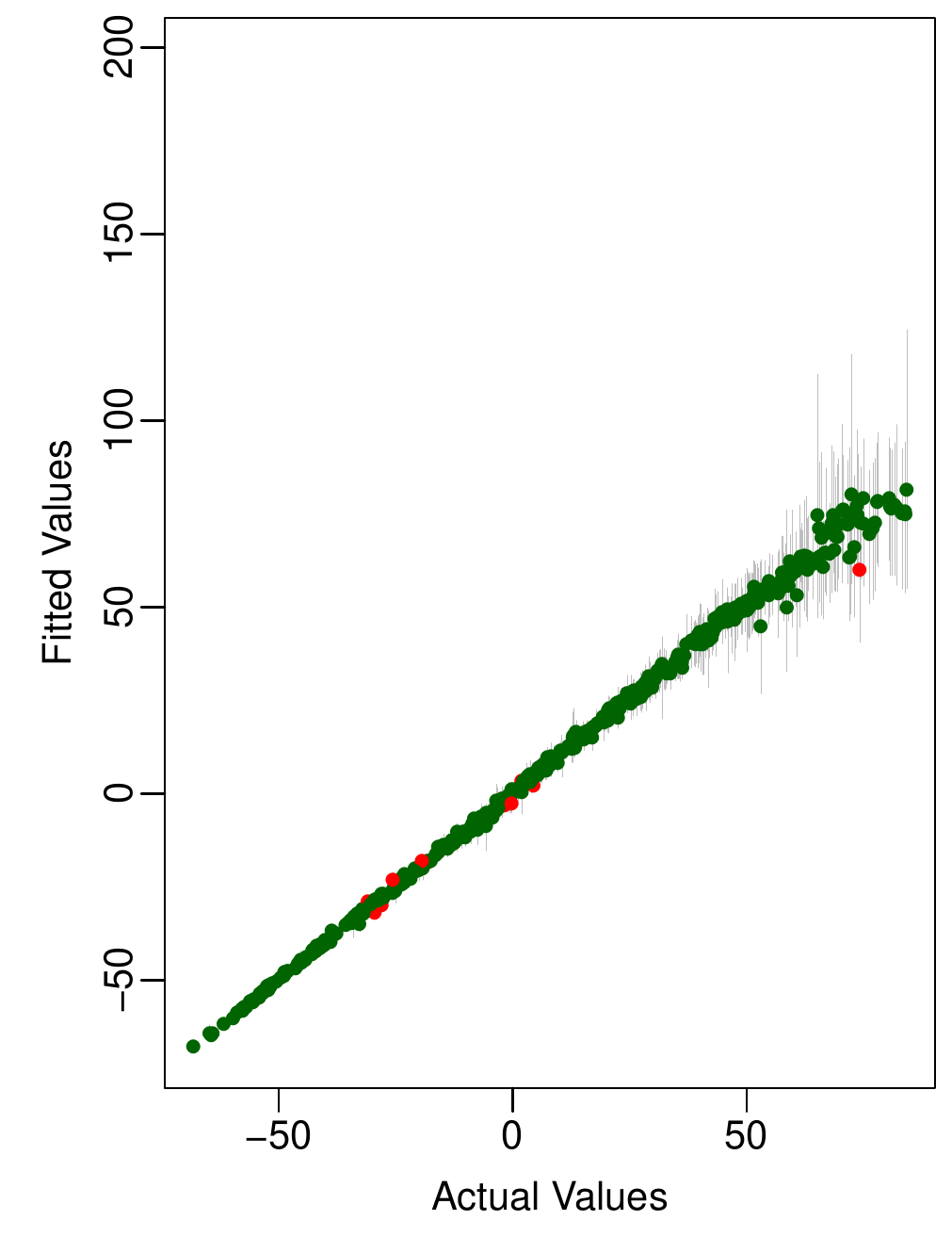}
\caption{\texttt{HBART}}
\label{subfig:cred_hetero}
\end{subfigure}~
\begin{subfigure}[b]{0.49\textwidth}
\centering
\includegraphics[width=3in]{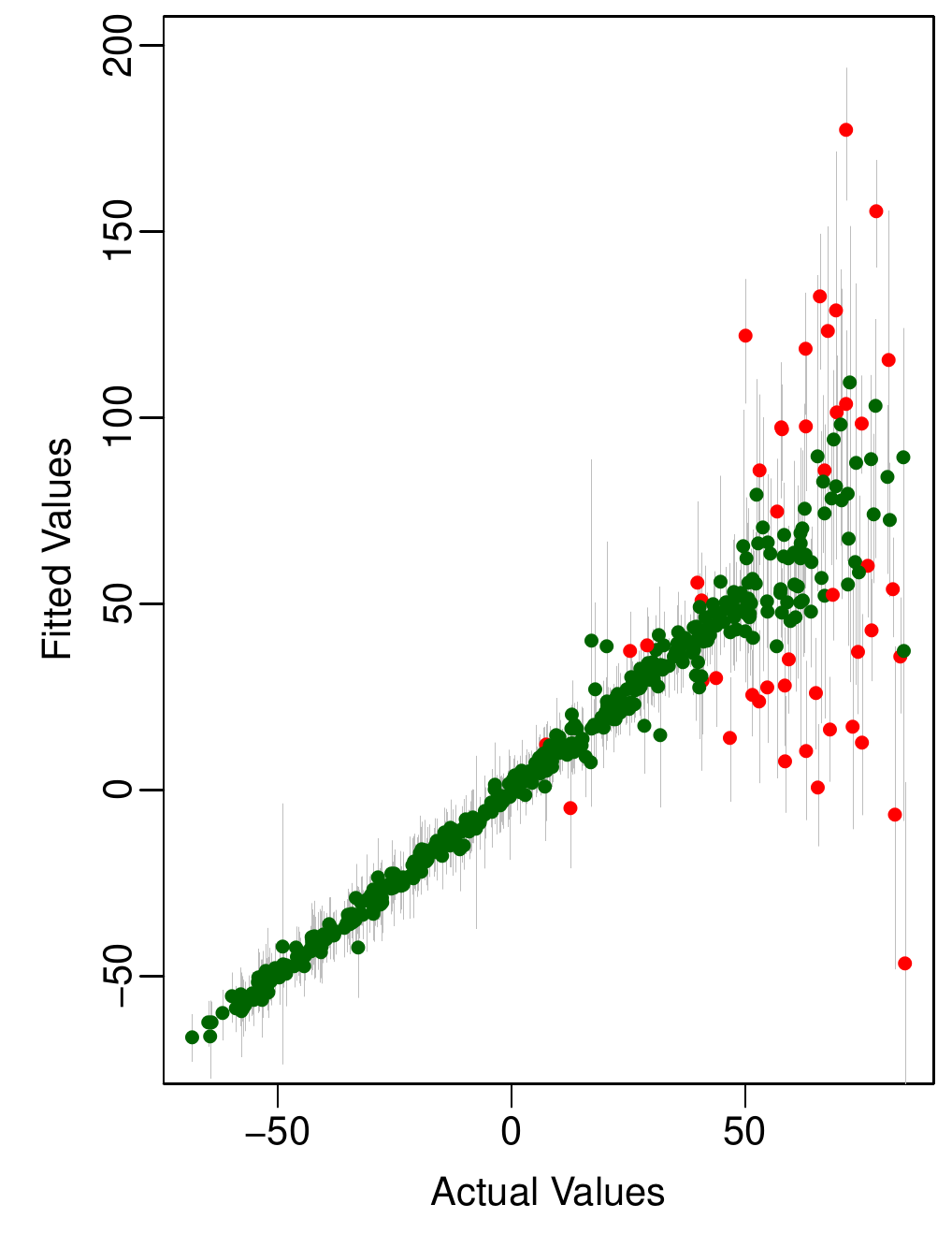}
\caption{\texttt{BART}}
\label{subfig:cred_homo}
\end{subfigure}
\caption{ Estimates of the conditional mean function $f$ for \texttt{HBART} and \texttt{BART} with associated 90\% credible intervals. The x-axis is the true value of the conditional mean function and the y-value is the model estimate. Gray lines illustrate 90\% credible intervals. If the true conditional mean falls within the interval, the point is colored \ingreen{green} and points in \inred{red} signify the true value falls outside of the interval.}
\label{fig:cred_intervals}
\end{figure}

Figure~\ref{fig:cred_intervals} plots actual values of the conditional mean function versus fitted values. This illustration demonstrates that in areas of high variance, \texttt{BART} has difficulty separating the mean function from the noise, and as a result, provides wide credible intervals for the true $f$. \texttt{HBART}, on the other hand, provides more narrow credible intervals, indicating that the algorithm was able to successfully separate the mean function from the heteroskedastic variance structure in the data generating process.

We next evaluate the performance of \texttt{HBART} versus \texttt{BART} in terms of out-of-sample predictive performance. We consider two models: one with homoskedastic errors, $\errorrv \iid \normnot{0}{3^2}$, and one with heteroskedastic error structure according to Equation~\ref{eq:multi_model_var}. For each error structure, we then generate $n=500$ training observations based on the data generating process outlined in Equation~\ref{eq:multi_model}. Both \texttt{HBART} and \texttt{BART} models are constructed on the training set and performance is evaluated in terms of root mean square error (RMSE) on an additional $n=500$ independent test observations drawn from the same data generating process. 

\begin{figure}[htp]
\centering
\begin{subfigure}[b]{0.49\textwidth}
\centering
\includegraphics[width=3in]{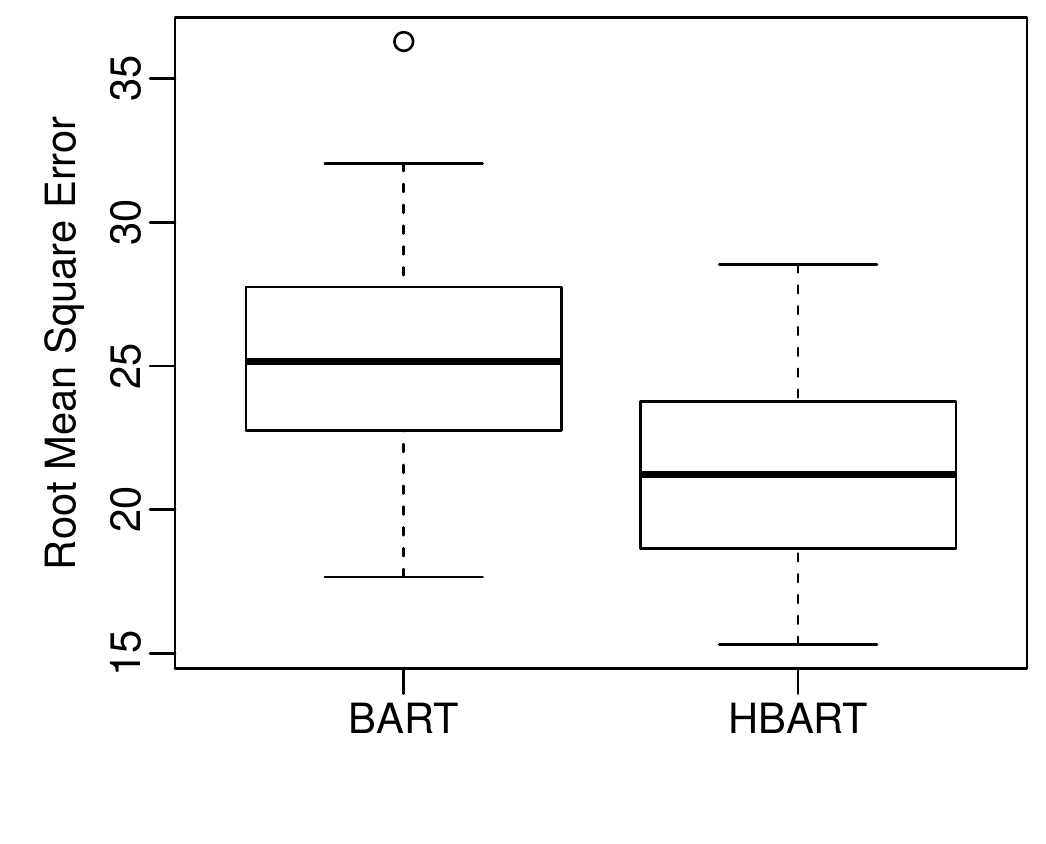}
\caption{Heteroskedastic Model}
\label{subfig:hetero_box}
\end{subfigure}~
\begin{subfigure}[b]{0.49\textwidth}
\centering
\includegraphics[width=3in]{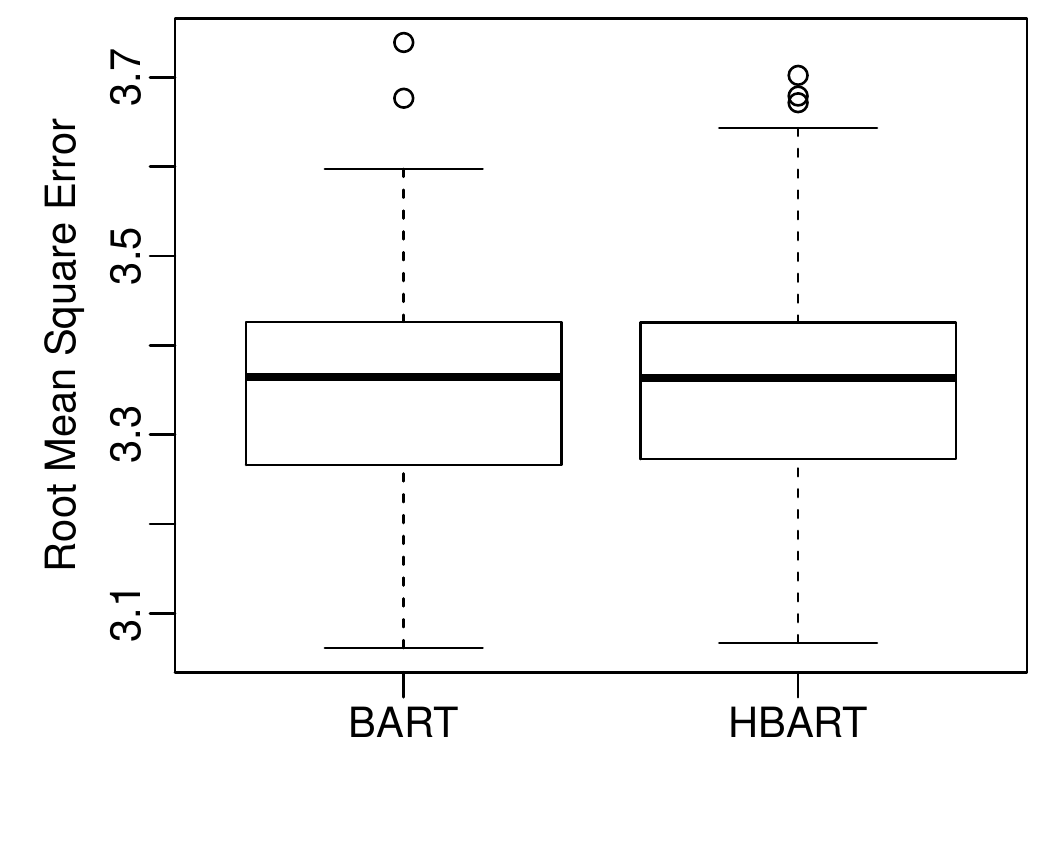}
\caption{Homoskedastic Model}
\label{subfig:homo_box}
\end{subfigure}
\caption{Distribution of out-of-sample RMSE for \texttt{BART} and \texttt{HBART} fit in 100 simulations for (a) the heteroskedastic data generating process of Equations~\ref{eq:multi_model} and \ref{eq:multi_model_var} and (b) the homoskedastic data generating process of Equation~\ref{eq:multi_model} with $\errorrv \iid \normnot{0}{3^2}$. The models are built on 500 observations and tested on 500 independent observations.}
\label{fig:rmse_boxplots}
\end{figure}

Figure~\ref{fig:rmse_boxplots} displays the out-of-sample performance results for 100 simulations. For the heteroskedastic model, \texttt{HBART} significantly outperforms \texttt{BART} (Figure~\ref{subfig:hetero_box}) For the homoskedastic model, the performance of \texttt{BART} and \texttt{HBART} are statistically equal and the results of both algorithms are quite stable (Figure~\ref{subfig:homo_box}). This plot provides more evidence of the robustness of \texttt{HBART}'s performance in the absence of heteroskedasticity.

\section{Real Data Examples}\label{sec:real_data}

\subsection{Lidar Data}

We consider the LIDAR data set explored in \citet{Ruppert2003}. The data set contains 221 observations. The response denoted \texttt{Log(Ratio)} is the logarithm of the ratio of reflected laser-emitted light from two sources and the predictor denoted \texttt{Range} is the distance traveled before the light is reflected back to it's source. \citet{Leslie2007} explored this data set by fitting both nonparametric mean and variance functions to the data under the assumptions of both normal and non-normal error distributions.

\begin{figure}[htp]
\centering
\begin{subfigure}[b]{0.49\textwidth}
\centering
\includegraphics[width=3in]{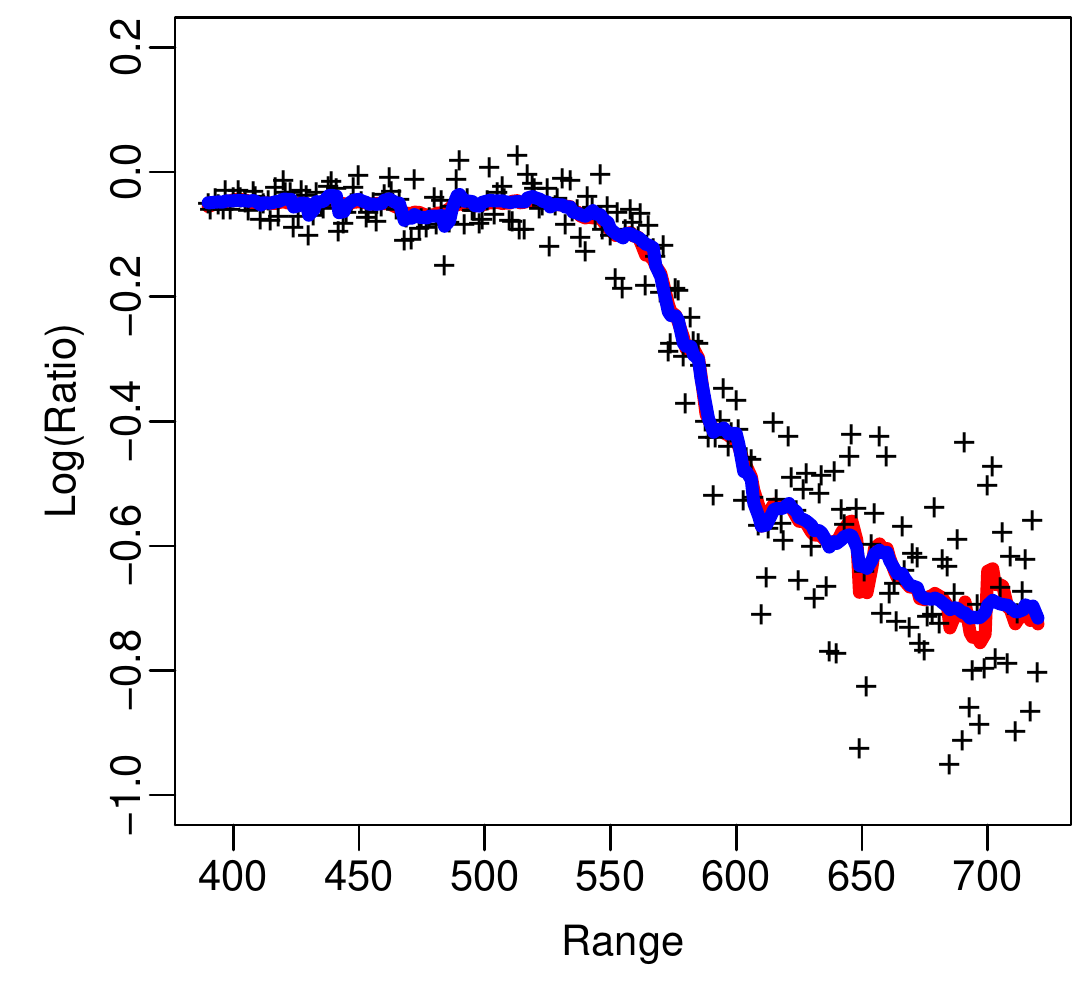}
\caption{Posterior Mean Estimate}
\label{subfig:lidar_data_mean_estimation}
\end{subfigure}~
\begin{subfigure}[b]{0.49\textwidth}
\centering
\includegraphics[width=3in]{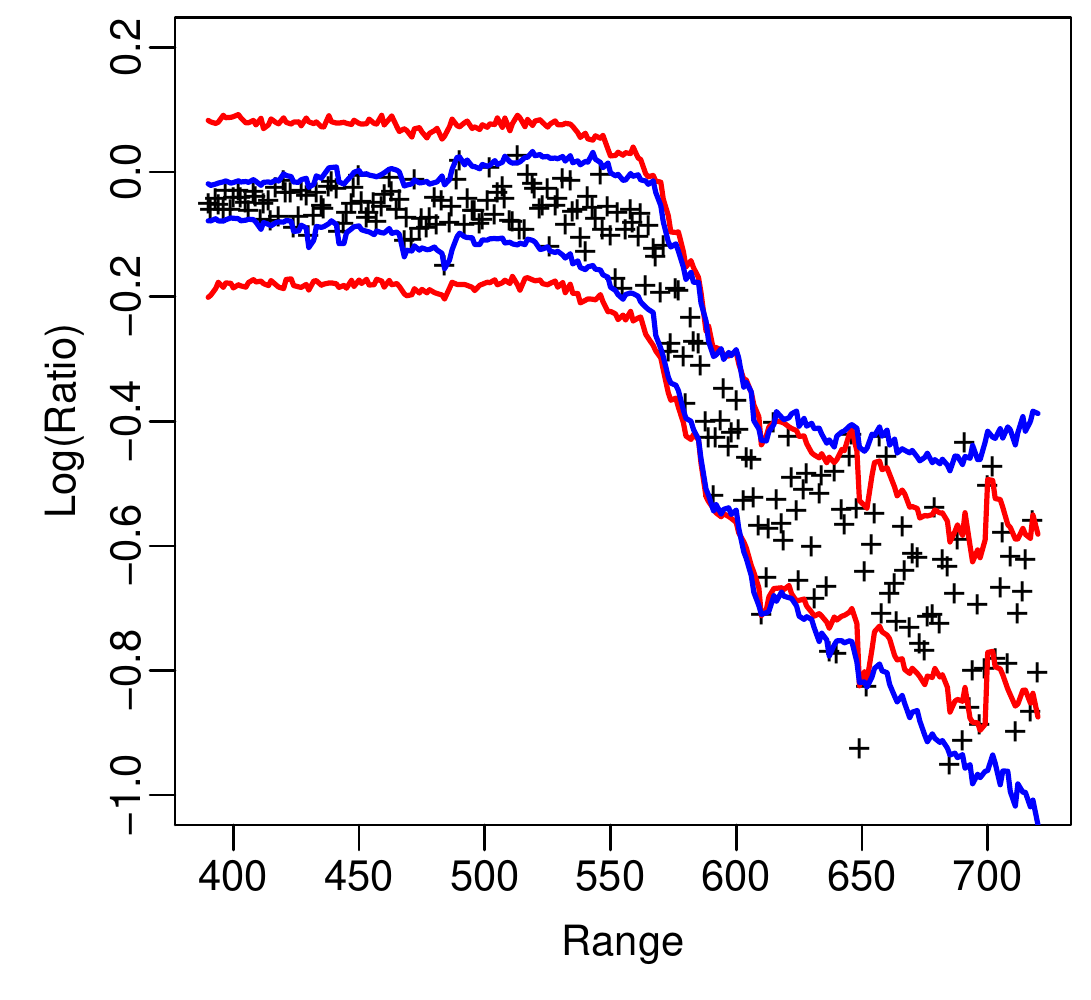}
\caption{Posterior Predictive Intervals}
\label{subfig:lidar_data_pred_intervals}
\end{subfigure}
\caption{\texttt{BART}'s and \texttt{HBART}'s posterior mean estimates and 90\% posterior predictive intervals for the Lidar data. \inred{Red} lines correspond to the \texttt{BART} model and \inblue{blue} lines correspond to the \texttt{HBART} model.}
\label{fig:lidar_data}
\end{figure}

We fit the data using \texttt{BART} and \texttt{HBART}. For \texttt{HBART}, $\Z$ is taken to be [\texttt{Range}, $\texttt{Range}^2$] where the two columns are orthogonalized. Figure~\ref{fig:lidar_data} illustrates the posterior mean estimates for both \texttt{HBART} and \texttt{BART}. One will notice that both algorithms estimate relatively similar posterior mean functions with \texttt{HBART}'s estimation being slightly more smooth in the region of higher variance than that of \texttt{BART}. Figure~\ref{subfig:pred_intervals} shows 90\% posterior predictive intervals for the two algorithms. The intervals for \texttt{HBART} seem to appropriately reflect the heteroskedasticity in the data, while \texttt{BART}'s predictive intervals are too wide at low values of \texttt{Range} and too narrow at high values of \texttt{Range}.

\subsection{Motorcycle Data}

We next consider a dataset of simulated motorcyle crashes that was compiled by \citet{Schmidt1981}. The observations in the data set consist of accelerometer readings (\texttt{acceleration}) taken from riders' helmets at 133 different points in time (\texttt{time}) after a simulated impact. As discussed in \citet{Gramacy2007}, many researchers find that this dataset exhibits multiple regimes in both the mean function and variance function over time. 

\begin{figure}[htp]
\centering
\includegraphics[width=4.2in]{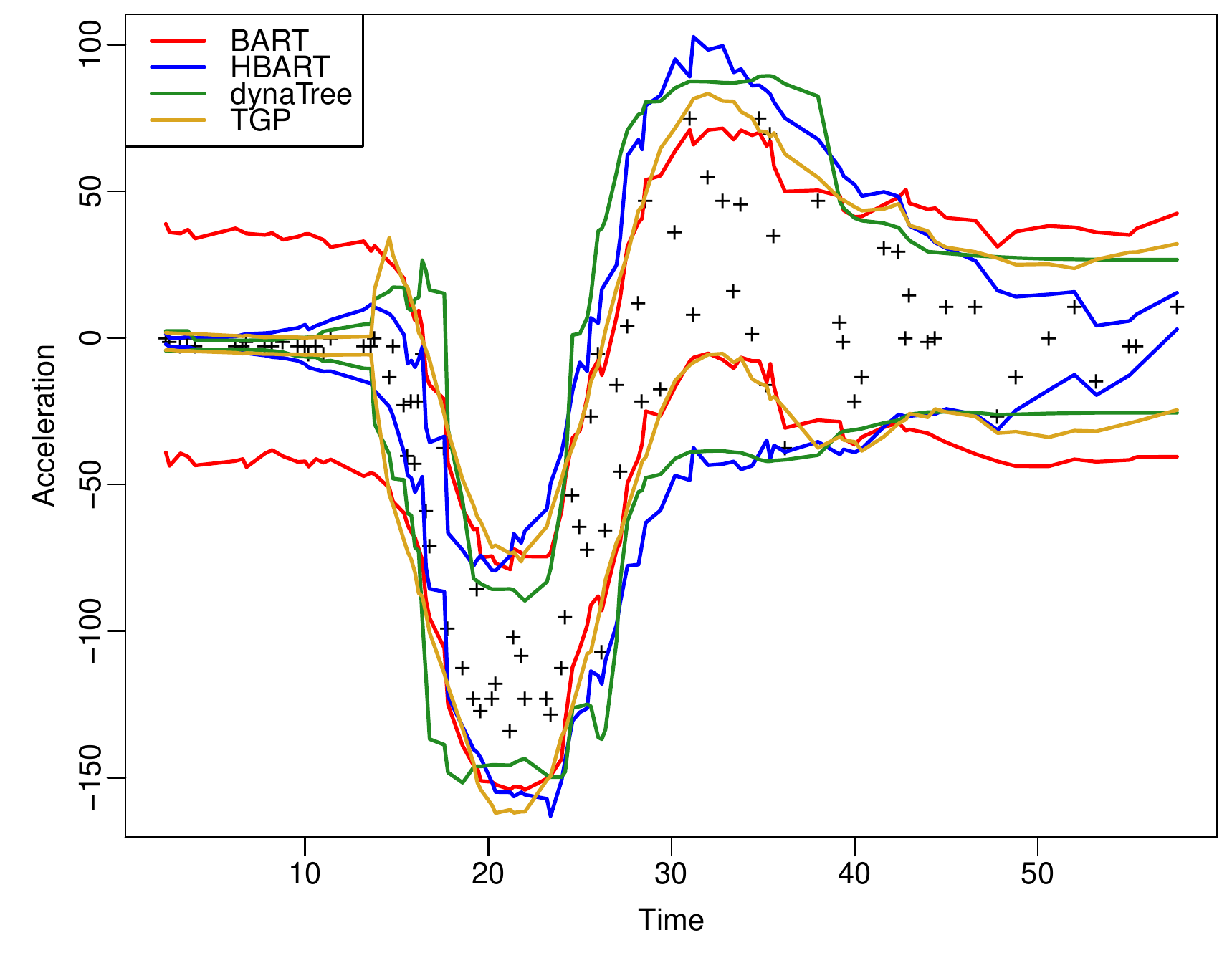}
\caption{90\% posterior predictive intervals for \texttt{BART}, \texttt{HBART}, \texttt{dynaTree} and \texttt{TGP}.}
\label{fig:motor}
\end{figure}

This dataset was also explored by \citet{Taddy2011} who remarked that the 90\% posterior predictive interval for \texttt{BART} appeared to \qu{variously over or under estimate data uncertainty around the regression mean. In particular, \texttt{BART}'s global variance term is misspecified for this heteroskedastic data.} We attempt to remedy this problem with \texttt{HBART}. Exploring a scatterplot of the data, the model seems to be characterized by a low variance regime followed by a high variance regime and then an additional low variance regime (see Figure~\ref{fig:motor}). Hence, when building an \texttt{HBART} model, we do not use the default $\Z$. To capture the low-high-low variance relationship, we specify the model quadratic in the predictor, $\Z = \bracks{\texttt{time},~\texttt{time}^2}$ where the two columns are orthogonalized. 

Figure~\ref{fig:motor} displays 90\% posterior predictive intervals for \texttt{HBART}, \texttt{BART}, dynamic trees \citep[\texttt{dynaTree},][]{Taddy2011}, and treed Gaussian processes \citep[\texttt{tgp},][]{Gramacy2007}. Each of the algorithms has a similar estimate of the posterior mean process (unshown), but there are some differences in the posterior predictive intervals. \texttt{BART}'s predictive intervals are much too wide at low and high values of time and perhaps too narrow for the intermediate values of time. However, \texttt{HBART}'s intervals are quite similar to those of \texttt{dynaTree}, being widest for intermediate values of time and more narrow near the beginning and end. Interestingly, \texttt{HBART} is the only of the four models that builds a narrower predictive interval at the higher values of time.

%\section{Parametric Models for Variance}\label{sec:parametric_models_for_variance}
%\inred{NUKE?}
%We present a way to go about building HBART models. We first run a vanilla BART. From here it would be interesting to ask the question: is there heteroskedasticity in this model at all? See Appendix~\ref{app:homoskedasticity_test} for our BART-based implementation of a test for homoskedasticity.
%
%If we find there is heteroskedasticity, we then build a BART model of the log squared residuals on the covariates and look at the partial dependence plots (PDPs) of each of the covariates. We then note the significant relationships we see (BART PDP's have 95\% credible intervals). We note whether these are linear or polynomial relationships and build an appropriate $Z$ matrix. We then can use a HBART model for prediction.
%
%We illustrate as an example in the XXX dataset. Here a BART model's residuals have a $p$ value of the test of homoskedasticity of 0. We now look at the PDP's of the p variables:\\
%
%[Figure with p subfigures]\\
%
%As we can see only 2 of these variables demonstrate something significant goin on. Variable j1 is linear and variable j2 looks like a quadratic. So our $Z$ matrix looks like $\bracks{x_{j1}, \tilde{x}_{j2}, \tilde{x}_{j2}^2}$ where the tilde notation is for orthogonal polynomials (not that this matters too much). 
%
%Now, doing a 10-fold build on 90\% and test on 90\% we approximate the oosRMSE for both BART and HBART and find the HBART does better because we properly modeled the RMSE. 

\section{Discussion}\label{sec:discussion}

We have proposed \texttt{HBART}, an extension of \texttt{BART} that relaxes the assumption of homoskedasticity in the model errors. In particular, we have developed a model that allows for a multiplicative heteroskedastic error structure, where the multiplicative factor is the exponential of a parametric function of some set of covariates. 

Through simulations and explorations of real data, we have demonstrated \texttt{HBART}'s potential for generating more appropriate posterior predictive intervals in the presence of appropriately modeled heteroskedastic data versus \texttt{BART}. Additionally, \texttt{BART} suffers from overfitting in areas of very high variance and \texttt{HBART} seems to offer promise in ameliorating this issue. In our explorations, the added complexity of fitting a model to the error terms did not hinder \texttt{HBART}'s performance on homoskedastic data. In this situation, \texttt{HBART}'s estimates of the posterior means, predictive intervals, and out-of-sample RMSE were very similar to \texttt{BART}'s.

Originally proposed in \citet{Chipman2010} and implemented in \citet{Kapelner2013}, it is possible to cross-validate over the a number of the hyperparameters of the \texttt{BART} model. Future work will involve extending \texttt{BART-CV} to \texttt{HBART-CV}, where it is possible to cross-validate over a selection of prior variances for the parametric variance model to impose varying degrees of shrinkage by modifying the $\Sigma_{jj}$ hyperparameters. One final direction of research is to further relax the assumptions on the error structure. For instance, one could incorporate more flexible variance models such as smoothing splines instead of the standard linear model or relax the assumption of normality of the errors by considering Dirichlet mixture priors.

\subsection*{Replication}

The code for \texttt{HBART} is located at \url{http://github.com/kapelner/bartMachine} in the branch \qu{hBART} and it is open source under the MIT license. The results, tables, and figures found in this paper can be replicated via the
\texttt{replication.R} script located in the \texttt{hbart\_paper} folder within this git repository. The code can be conveniently installed as a local \texttt{R} package. 

\subsection*{Acknowledgements}

We thank Richard Berk, Ed George and Abba Krieger for helpful discussions. We thank Simon Urbanek for his generous help with rJava. Adam Kapelner acknowledges support from the National Science Foundation's Graduate Research Fellowship Program.

\bibliographystyle{apalike}
\bibliography{working_paper}

\begin{thebibliography}{}

\bibitem[Boscardin and Gelman, 1994]{Boscardin1994}
Boscardin, W. and Gelman, A. (1994).
\newblock {Bayesian Computation for Parametric Models of Heteroscedasticity in
  the Linear Model}.
\newblock {\em University of California, Berkeley}.

\bibitem[Carroll and Ruppert, 1988]{Carroll1988}
Carroll, R. and Ruppert, D. (1988).
\newblock {\em Transformation and Weighting in Regression}.
\newblock Chapman \& Hall/CRC Monographs on Statistics \& Applied Probability.
  Taylor \& Francis.

\bibitem[Cepeda and Gamerman, 2001]{Cepeda2001}
Cepeda, E. and Gamerman, D. (2001).
\newblock {Bayesian modeling of variance heterogeneity in normal regression
  models}.
\newblock {\em Brazilian Journal of Probability and Statistics}, pages
  207--221.

\bibitem[Chan et~al., 2006]{Chan2006}
Chan, D., Kohn, R., Nott, D., and Kirby, C. (2006).
\newblock {Locally adaptive semiparametric estimation of the mean and variance
  functions in regression models}.
\newblock {\em Journal of Computational and Graphical Statistics}, pages 0--26.

\bibitem[Chipman et~al., 2010]{Chipman2010}
Chipman, H., George, E., and McCulloch, R. (2010).
\newblock {BART: Bayesian Additive Regressive Trees}.
\newblock {\em The Annals of Applied Statistics}, 4(1):266--298.

\bibitem[Gamerman, 1997]{Gamerman1997}
Gamerman, D. (1997).
\newblock {Sampling from the posterior distribution in generalized linear mixed
  models}.
\newblock {\em Statistics and Computing}.

\bibitem[Gelman et~al., 2004]{Gelman2004}
Gelman, A., Carlin, J., Stern, H., and Rubin, D. (2004).
\newblock {\em Bayesian Data Analysis}.
\newblock Chapman \& Hall / CRC, second edition.

\bibitem[Geman and Geman, 1984]{Geman1984}
Geman, S. and Geman, D. (1984).
\newblock Stochastic relaxation, {G}ibbs distributions, and the {B}ayesian
  restoration of images.
\newblock {\em IEEE Transaction on Pattern Analysis and Machine Intelligence},
  6:721--741.

\bibitem[Gramacy, 2007]{Gramacy2007}
Gramacy, R.~B. (2007).
\newblock {tgp: An R Package for Bayesian Nonstationary, Semiparametric
  Nonlinear Regression and Design by Treed Gaussian Process Models}.
\newblock {\em Journal Of Statistical Software}, 19(9).

\bibitem[Hastie and Tibshirani, 2000]{Hastie2000}
Hastie, T. and Tibshirani, R. (2000).
\newblock {Bayesian Backfitting}.
\newblock {\em Statistical Science}, 15(3):196--213.

\bibitem[Hastings, 1970]{Hastings1970}
Hastings, W. (1970).
\newblock {Monte Carlo sampling methods using Markov chains and their
  applications}.
\newblock {\em Biometrika}, 57(1):97--109.

\bibitem[Kapelner and Bleich, 2013]{Kapelner2013}
Kapelner, A. and Bleich, J. (2013).
\newblock bartmachine: Machine learning with bayesian additive regression
  trees.
\newblock {\em ArXiv}.

\bibitem[Leslie et~al., 2007]{Leslie2007}
Leslie, D., Kohn, R., and Nott, D. (2007).
\newblock {A general approach to heteroscedastic linear regression}.
\newblock {\em Statistics and Computing}, pages 1--29.

\bibitem[Ruppert et~al., 2003]{Ruppert2003}
Ruppert, D., Wand, P., and Carroll, R. (2003).
\newblock {\em Semiparametric Regression}.
\newblock Cambridge Series in Statistical and Probabilistic Mathematics.
  Cambridge University Press.

\bibitem[Schmidt et~al., 1981]{Schmidt1981}
Schmidt, G., Mattern, R., and Sch{\"u}ler, F. (1981).
\newblock Biomechanical investigation to determine physical and traumatological
  differentiation criteria for the maximum load capacity of head and vertebral
  column with and without protective helmet under effect.
\newblock Technical report, EFC Research Program on Biomechanics of Impacts.
  Final report Phase III, Project G5. Institute f{\"u}r Rechtsmedizin,
  Universit{\"a}t Heidelberg.

\bibitem[Taddy et~al., 2011]{Taddy2011}
Taddy, M., Gramacy, R., and Polson, N. (2011).
\newblock {Dynamic Trees for Learning and Design}.
\newblock {\em Journal of the American Statistical Association},
  106(493):109--123.

\bibitem[White, 1980]{White1980}
White, H. (1980).
\newblock A heteroskedasticity-consistent covariance matrix estimator and a
  direct test for heteroskedasticity.
\newblock {\em Econometrica}, pages 817--838.

\bibitem[Yau and Kohn, 2003]{Yau2003}
Yau, P. and Kohn, R. (2003).
\newblock {Estimation and variable selection in nonparametric heteroscedastic
  regression}.
\newblock {\em Statistics and Computing}, pages 191--208.

\end{thebibliography}

\appendix
\section{Gibbs Sampling Details for \texttt{HBART}}\label{app:details}

In the following sections, we provide implementation details for each step of the \texttt{HBART} Gibbs sampler (Equation~\ref{eq:gibbs_sampler}).

%\bneqn \label{eq:gibbs_sampler}
%1: & \treet{1} &|~~ R_{-1}, \sigsq, \bgamma \\
%2:& \leaf_1 &|~~ \treet{1}, R_{-1}, \sigsq, \bgamma \nonumber\\ 
%&\vdots \nonumber\\
%2m - 1:& \treet{m} &|~~ R_{-m}, \sigsq, \bgamma\nonumber\\
%2m:& \leaf_m &|~~ \treet{m}, R_{-m}, \sigsq, \bgamma \nonumber\\ 
%2m + 1:& \sigsq &|~~ \treeleaft{1}, \ldots, \treeleaft{m}, \bgamma, \E \nonumber\\
%2m + 2:& \bgamma &|~~ \sigsq, \nonumber\E
%\eneqn

\subsection*{Drawing $\sigsq$ (step $2m+1$)}

Drawing $\sigsq$ in \texttt{HBART} requires a slight modification to the scheme used in the original homoskedastic \texttt{BART}. With the errors distributed multivariate normal (Equation~\ref{eq:hetero_bart_model}) and the prior on $\sigsq$ being distributed inverse-gamma, conjugacy yields the posterior as an inverse-gamma distribution just as in \texttt{BART}. The quadratic form of the errors and their covariance matrix carries into the scale parameter. Hence, we sample $\sigsq$ as

\beqn
\sigsq &\sim& \invgammanot{\overtwo{\nu + n}}{\overtwo{\nu\lambda + \berrorrv^T\D^{-1}\berrorrv}} \\
&=& \invgammanot{\overtwo{\nu + n}}{\overtwo{\nu\lambda} + \half\parens{ \sum_{i=1}^n \exp{-\z_i^\top \bgamma}\errorrv_i^2}}.
\eeqn

\noindent The default values of hyperparameters $\nu$ and $q$ are set to be the same as the defaults of the original homoskedastic \texttt{BART} algorithm.

\subsection*{Drawing $\bgamma$ (step $2m+2$)}

The posterior of $\bgamma ~|~ \errorrv_1, \ldots, \errorrv_n, \sigsq$ is proportional to:

\small
\beqn
\cprob{\bgamma}{\errorrv_1, \ldots, \errorrv_n, \sigsq} &\propto& \cprob{\errorrv_1, \ldots, \errorrv_n}{\bgamma, \sigsq} \prob{\bgamma} \\
&=& \prod_{i=1}^n \normpdfmeanzero{\errorrv_i}{\sigsq\exp{\z_i^\top \bgamma}} \times \\
&& \prod_{j=1}^p \normpdf{\gamma_j}{\gamma_{0j}}{\Sigma_{jj}} \\
&\propto& \oneoversqrt{\sigsq\exp{\sum_{i=1}^n \z_i^\top \bgamma}}\exp{-\half \parens{\sum_{i=1}^n \frac{\errorrv_i^2}{{\sigsq\exp{\z_i^\top \bgamma}}} + \sum_{j=1}^p \frac{\squared{\gamma_j - \gamma_{0j}}}{\Sigma_{jj}}}}
\eeqn
\normalsize

%The log probability is proportional to:
%
%\inblue{
%\beqn
%-\half \parens{\natlog{\sigsq} + \sum_{i=1}^n \z_i^\top \bgamma + \sum_{i=1}^n \frac{\errorrv_i^2}{\sigsq\exp{\z_i^\top \bgamma}} + \sum_{j=1}^p \frac{\squared{\gamma_j - \gamma_{0j}}}{\Sigma_{jj}}}
%\eeqn
%}

\noindent Since it is not tractable to draw directly from the above distribution, a proposal distribution is required. A suitable proposal distribution can be obtained using \citet{Gamerman1997}, which is based on a single step of an iteratively reweighted least squares algorithm. Letting $\bgamma$ represent the current value of the parameter vector, the recommended proposal density is $\prob{\bgamma^*|\bgamma } = \multnormnot{p}{\bv{a}(\bgamma)}{\B}$ where 

\small
\beqn
&& \bv{a} = \B \parens{\bSigmainv \bgamma_0 + \half \Z^T \bv{w}}, \quad \B = \inverse{\bSigmainv + \half\Z^T\Z} \text{and} \\ 
&& \bv{w} = \bracks{\z_1^\top \bgamma + \frac{\errorrv_1^2}{\sigsq\exp{\z_1^\top \bgamma}}-1 \ldots \z_n^\top \bgamma + \frac{\errorrv_n^2}{\sigsq\exp{\z_n^\top \bgamma}}-1 }^\top. \\
\eeqn
\normalsize

\noindent Then, $\bgamma^*$ is accepted if a draw from a standard uniform distribution is less than the Metropolis-Hastings ratio \citep[p.291]{Gelman2004},

\bneqn\label{eq:mh_step_gamma}
r = \underbrace{\frac{\prob{\bgamma^* | \bgamma}}{\prob{\bgamma | \bgamma^*}}}_{\text{jump ratio}} \underbrace{\frac{\cprob{\gamma^*}{\sigsq, \errorrv_1, \ldots, \errorrv_n}}{\cprob{\gamma}{\sigsq, \errorrv_1, \ldots, \errorrv_n}}}_{\text{likelihood ratio}}.
\eneqn

\noindent The \textit{jump ratio} in Equation~\ref{eq:mh_step_gamma} becomes

\beqn
\frac{\prob{\bgamma^* |\bgamma}}{\prob{\bgamma |\bgamma^*}} &=& \frac{{\tothepow{2\pi}{-\overtwo{p}}} {\abss{\B}^{-\half}} \exp{-\half \transp{\bgamma^* - a(\bgamma)}\B^{-1} \parens{\bgamma^* - a(\bgamma)}}}{{\tothepow{2\pi}{-\overtwo{p}}} {\abss{\B}^{-\half}} \exp{-\half \transp{\bgamma - a(\bgamma^*)}\B^{-1} \parens{\bgamma - a(\bgamma^*)}}} \\
&=& \exp{\half{\transp{\bgamma - \bgamma^* + a(\bgamma) - a(\bgamma^*)} \B^{-1} \parens{\bgamma - \bgamma^* + a(\bgamma) - a(\bgamma^*)}}},
\eeqn

\noindent the \textit{likelihood ratio} in Equation~\ref{eq:mh_step_gamma} is calculated to be

\small
\beqn
\frac{\cprob{\gamma^*}{\sigsq, \errorrv_1, \ldots, \errorrv_n}}{\cprob{\gamma}{\sigsq, \errorrv_1, \ldots, \errorrv_n}} &=& \frac{\displaystyle \prod_{i=1}^n \normpdfmeanzero{\errorrv_i}{\sigsq\exp{\z_i^\top \bgamma^*}}}{\displaystyle\prod_{i=1}^n \normpdfmeanzero{\errorrv_i}{\sigsq\exp{\z_i^\top \bgamma}}} \\
&& \times ~\frac{\displaystyle\prod_{j=1}^p \normpdf{\gamma_j^*}{\gamma_{0j}}{\Sigma_{jj}}}{\displaystyle\prod_{j=1}^p \normpdf{\gamma_j}{\gamma_{0j}}{\Sigma_{jj}}}\\
%&=& \parens{\prod_{i=1}^n \sqrt{\frac{\exp{\z_i^\top \bgamma}}{\exp{\z_i^\top \bgamma^*}}}}\exp{-\oneover{2\sigsq} \parens{\sum_{i=1}^n \frac{\errorrv_i^2}{\exp{\z_i^\top \bgamma^*}} + \frac{\errorrv_i^2}{\exp{\z_i^\top \bgamma}}}} \exp{\half\sum_{j=1}^p \oneover{\Sigma_{jj}}\parens{\squared{\gamma_j - \gamma_{0j}} - \squared{\gamma_j^* - \gamma_{0j}}}}\\
%&=& {{\exp{\half \sum_{i=1}^n \z_i^\top (\bgamma - \bgamma^*)}}}\exp{\oneover{2\sigsq} \parens{\sum_{i=1}^n \errorrv_i^2 \parens{\exp{-\z_i^\top \bgamma} - \exp{-\z_i^\top \bgamma^*} }}} \exp{\half\sum_{j=1}^p \oneover{\Sigma_{jj}}\parens{\gamma_j^2 - \gamma_j^{*^2} + 2\gamma_{0j}(\gamma_j^* - \gamma_j)}} \\
&=& \text{exp}\Bigg(\half \Bigg(\sum_{i=1}^n \z_i^\top (\bgamma - \bgamma^*) \\
&&+~ \oneover{\sigsq} \parens{\sum_{i=1}^n \errorrv_i^2 \parens{\exp{-\z_i^\top \bgamma} - \exp{-\z_i^\top \bgamma^*}}\Bigg)} \\
&&+~ \sum_{j=1}^p \oneover{\Sigma_{jj}}\parens{\gamma_j^2 - \gamma_j^{*^2} + 2\gamma_{0j}(\gamma_j^* - \gamma_j)}\Bigg)
\eeqn
\normalsize

\noindent which results in the Metropolis-Hastings ratio of Equation~\ref{eq:mh_step_gamma} of

\small
\beqn
r &=& \text{exp}\Bigg(\half \Bigl(\transp{\bgamma - \bgamma^* + a(\bgamma) - a(\bgamma^*)} \B^{-1} \parens{\bgamma + \bgamma^* - \parens{a(\bgamma) + a(\bgamma^*)}} \\
&& +~ \sum_{i=1}^n \z_i^\top (\bgamma - \bgamma^*) + \oneover{\sigsq}\sum_{i=1}^n \errorrv_i^2 \parens{\exp{-\z_i^\top \bgamma} - \exp{-\z_i^\top \bgamma^*}} \\
&& +~ \sum_{j=1}^p \oneover{\Sigma_{jj}} \parens{\gamma_j^2 - \gamma_j^{*2} + 2\gamma_{0j} \parens{\gamma_j^* - \gamma_j}} \Bigr)\Bigg).
\eeqn
\normalsize

We now choose default values for the hyperparameters $\gamma_{0,1}, \ldots, \gamma_{0,p}$ and $\Sigma_{11}, \ldots, \Sigma_{pp}$. All $\gamma_{0,j}$ are set to 0. This choice centers the prior distribution of the linear model coefficients at zero. For the variance hyperparameters, we choose the $\Sigma_{jj}$'s to be 1000 for all $j$ which is sufficiently large so our model will not impose shrinkage of the coefficients towards 0. Investigating this parameter's role in our algorithm we view as fruitful future work.

\subsection*{Drawing the $\leaf$'s (steps $2,4, \ldots, 2m$)}

Sampling the leaf parameters must be adjusted to reflect the heteroskedasticity in the model. Observations considered highly variable will now \textit{be downweighted} when constructing an estimate of a terminal node's prediction. Recall that $\leaf_t ~|~ \treet{1}, \R_1, \sigsq_1, \ldots, \sigsq_n$ is the sampling for all leaves where each leaf is considered independent,

\beqn
\mu_{t1} &~|~& \treet{t}, \R_{t_1}, \sigsq_1, \ldots, \sigsq_n \\
\mu_{t2} &~|~& \treet{t}, \R_{t_2}, \sigsq_1, \ldots, \sigsq_n \\
\vdots && \\
\mu_{tb_t} &~|~& \treet{t}, \R_{t_{b_t}}, \sigsq_1, \ldots, \sigsq_n
\eeqn

\noindent where $b_t$ denotes the number of terminal nodes in the $t$th tree and the subscripts on the $R_{t_i}$ terms denote only the data that is apportioned to the specific terminal node. Recall that the prior on the leaf parameters, $\mu$'s, are normal and the likelihood of the responses, $R_{t}$'s, are assumed normal as well. 

Given the normal-normal conjugacy, we derive the normal posterior distribution for a given leaf's prediction which is necessarily a function of the heterogeneous variances. We drop the subscripts on the $R$ term for convenience and denote $k$ as the number of data records that fell into the given leaf. Note that we drop double subscripting on the variances $\braces{\sigsq_1, \ldots, \sigsq_k}$ which are a subset of $\braces{\sigsq_1, \ldots, \sigsq_n}$ for the data records in this leaf. 

\beqn
\cprob{\mu}{\R, \sigsq_1, \ldots, \sigsq_k, \bgamma} &\propto& \cprob{R}{\mu, \sigsq_1, \ldots, \sigsq_k, \bgamma} \cprob{\mu}{\sigsq_1, \ldots, \sigsq_k} \\
&=& \multnormnot{k}{\mu\onevec}{\sigsq\D} \normnot{0}{\sigsq_\mu} = \parens{\prod_{i=1}^{k}\normnot{\mu}{\sigsq_i}}\normnot{0}{\sigsq_\mu} \\
%&=& \normnot{\inverse{\oneover{\sigsqmu} \I_k + k \D^{-1}} \parens{k\D^{-1}\Rbar\onevec}}{\inverse{\oneover{\sigsqmu} \I_k + k \D^{-1}}} \\
&=& \normnot{\frac{\displaystyle \sum_{i=1}^k\dfrac{R_i}{\sigsq_i}}{\doneover{\sigsq_\mu} + \displaystyle\sum_{i=1}^k \dfrac{1}{\sigsq_i}}}{\oneover{\doneover{\sigsq_\mu} + \displaystyle\sum_{i=1}^k \dfrac{1}{\sigsq_i}}}
\eeqn

%\subsection*{Changes in Gibbs Sampler for $\sigsq_i$}
%
%In vanilla BART, the posterior of $\sigsq$ was:
%
%\beqn
%\sigsq ~|~ e_1, \ldots, e_n \sim \invgammanot{\overtwo{\nu + n}}{\overtwo{\lambda\nu + \sum_{i=1}^n e_i^2}} 
%\eeqn
%
%Now, we have to sample each $\sigsq_1, \ldots, \sigsq_n$ from their individual posteriors. The analagous expression is tha t the SSE is based on only one residual:
%
%\beqn
%\sigsq_i ~|~ e_i \sim \invgammanot{\overtwo{\nu + 1}}{\overtwo{\lambda\nu + e_i^2}}
%\eeqn

\subsection*{Drawing the $\treet{}$'s (steps $1, 3, \ldots, 2m - 1$)}

As described in \citet[Appendix A]{Kapelner2013}, the draw of a new tree structure relies on a Metropolis-Hastings step where trees can be altered via growing new daughter nodes from an existing terminal node (\texttt{GROW}), pruning two terminal nodes such that their parent becomes terminal (\texttt{PRUNE}), or changing the splitting rule in a node (\texttt{CHANGE}). 

Below is the Metropolis-Hastings ratio where the parameter sampled is the tree and the data is the responses unexplained by other trees denoted by $\R$. We denote the new, proposal tree with an asterisk and the original tree without the asterisk.

\bneqn\label{eq:naive_metropolis}
r = \frac{\prob{\treet{*} \rightarrow \treet{}}}{\prob{\treet{} \rightarrow \treet{*}}} \frac{\cprob{\treet{*}}{\R, \sigsq, \bgamma}}{\cprob{\treet{}}{\R, \sigsq, \bgamma}}
\eneqn

This can be reformulated using repeated applications of Bayes Rule to be a product of three ratios.

\beqn
r &=& \underbrace{\frac{\prob{\treet{*} \rightarrow \treet{}}}{\prob{\treet{} \rightarrow \treet{*}}}}_{\text{transition ratio}} ~~~\times~~~ \underbrace{\frac{\cprob{\R}{\treet{*}, \sigsq, \bgamma}}{\cprob{\R}{\treet{}, \sigsq, \bgamma}}}_{\text{likelihood ratio}} ~~~\times \underbrace{\frac{\prob{\treet{*}}}{\prob{\treet{}}}}_{\text{tree structure ratio}}
\eeqn

\noindent Note that the probability of the tree structure is independent of $\sigsq$ and $\bgamma$. 

The transition ratio and the tree structure ratio remain the same as in the original \texttt{BART} as they do not depend on the variance parameters. The likelihood ratio now must take into account the heterogeneity in variances. The \texttt{PRUNE} likelihood ratio is the inverse of the \texttt{GROW} likelihood ratio. Thus, we only need to compute likelihood ratios for \texttt{GROW} and \texttt{CHANGE}. 

\subsubsection*{The Likelihood Ratio for the \texttt{GROW} proposal}

Since the likelihoods are solely determined by the terminal nodes, the grown proposal tree differs from the original tree by only the selected node to be grown. We denote the node to be grown by $\ell$, the left child by $\ell_L$ and the right child by $\ell_R$.

\small
\bneqn\label{eq:grow_lik_ratio}
\frac{\cprob{\R}{\treet{*} \sigsq, \bgamma}}{\cprob{\R}{\treet{}, \sigsq, \bgamma}} &=& \frac{\displaystyle\myint{\mu}{\reals}{}{\cprob{\R_{\ell_L}}{\mu, \sigsq, \bgamma}\prob{\mu}} ~~ \displaystyle\myint{\mu}{\reals}{}{\cprob{\R_{\ell_R}}{\mu, \sigsq, \bgamma}\prob{\mu}}}{\displaystyle\myint{\mu}{\reals}{}{\cprob{\R_\ell}{\mu, \sigsq, \bgamma}\prob{\mu}}} 
\eneqn
\normalsize

Each of these three integrals are the same with regards to marginalizing the $\mu$ term:

\bneqn\label{eq:marginalization}
&& \myint{\mu}{\reals}{}{\cprob{\R}{\mu, \sigsq, \bgamma}\prob{\mu}} \nonumber \\
&=& \myint{\mu}{\reals}{}{\oneover{\tothepow{2\pi}{\overtwo{n}} \abss{\D}^{-\half}}\exp{-\half\transpose{\R - \mu\onevec} \D^{-1} (\R - \mu\onevec)} ~\oneoversqrt{2\pi\sigsq_\mu} \exp{-\oneover{2\sigsq_\mu} \musq}}~ \nonumber \\
%&=& \myint{\mu}{\reals}{}{\parens{\prod_{i=1}^{n} \oneoversqrt{2\pi\sigsq_i} \exp{-\oneover{2\sigsq_i} \squared{R_i - \mu}}} ~\oneoversqrt{2\pi\sigsq_\mu} \exp{-\oneover{2\sigsq_\mu} \musq}}~ \\
%&=& \oneover{\tothepow{2\pi}{\overtwo{n} + 1}\sqrt{\sigsqmu \prod_{i=1}^{n} \sigsq_i}}\myint{\mu}{\reals}{}{\exp{-\half \parens{\musq + \sumonen{i}{\squared{R_i - \mu}}} }}\\
%&=& \oneover{\tothepow{2\pi}{\overtwo{n} + 1}\sqrt{\sigsqmu \prod_{i=1}^{n} \sigsq_i}}\myint{\mu}{\reals}{}{\exp{-\half \parens{\frac{\musq}{\sigsq_\mu} + \sumonen{i}{\frac{R_i^2}{\sigsq_i} - \frac{2\mu R_i}{\sigsq_i} + \frac{\musq}{\sigsq_i}}} }}\\
%&=& \oneover{\tothepow{2\pi}{\overtwo{n} + 1}\sqrt{\sigsqmu \prod_{i=1}^{n} \sigsq_i}}\myint{\mu}{\reals}{}{\exp{-\half \parens{\frac{\musq}{\sigsq_\mu} + \sumonen{i}{\frac{R_i^2}{\sigsq_i}} - 2\mu \sumonen{i}{\frac{R_i}}{\sigsq_i} + \musq\sumonen{i}{\frac{1}{\sigsq_i}}} }}\\
&=& \oneover{\tothepow{2\pi}{\overtwo{n} + 1}\sqrt{\sigsqmu \prod_{i=1}^{n} \sigsq_i}} \exp{-\half \sumonen{i}{\frac{R_i^2}{\sigsq_i}}} \nonumber \\
&& \times ~\myint{\mu}{\reals}{}{\exp{-\half \parens{\frac{\musq}{\sigsq_\mu} - 2\mu \sumonen{i}{\frac{R_i}}{\sigsq_i} + \musq\sumonen{i}{\frac{1}{\sigsq_i}}} }} \nonumber\\
%\eeqn
%
%Now we can solve the underbraced integral above using Mathematica:
%
%%\begin{figure}[htp]
%%\centering
%%\includegraphics[width= 5.0in]{mathematica.jpg}
%%\end{figure}
%
%Plugging this result into the equation where the integral once was, we obtain:
%
%\beqn
%&=& \oneover{\tothepow{2\pi}{\overtwo{n} + 1}\sqrt{\sigsqmu \prod_{i=1}^{n} \sigsq_i}} \exp{-\half \sumonen{i}{\frac{R_i^2}{\sigsq_i}}} \sqrt{\frac{2\pi}{\oneover{\sigsq_\mu} + \displaystyle \sum_{i=1}^n \oneover{\sigsq_i}}} \exp{\frac{\sigsq_\mu \squared{\displaystyle \sum_{i=1}^n \frac{R_i}{\sigsq_i}}}{2 + 2\sigsq_\mu \displaystyle \sum_{i=1}^n \oneover{\sigsq_i}}}\\
&=& {\tothepow{(2\pi)^n \parens{1 + \sigsq_\mu \displaystyle \sum_{i=1}^{n} \oneover{\sigsq_i}} \displaystyle \prod_{i=1}^{n} \sigsq_i}{-\half}}~ \exp{\half \parens{\frac{\sigsq_\mu\squared{\displaystyle \sum_{i=1}^n \frac{R_i}{\sigsq_i}}}{1 + \sigsq_\mu \displaystyle \sum_{i=1}^n \oneover{\sigsq_i}}- \sumonen{i}{\frac{R_i^2}{\sigsq_i}}}}
\eneqn

\noindent The likelihood ratio can now be computed by substituting Equation~\ref{eq:marginalization} into Equation~\ref{eq:grow_lik_ratio} three times to arrive at:

%\beqn
%\frac{\cprob{\R}{\treet{*}, \sigsq, \bgamma}}{\cprob{\R}{\treet{}, \sigsq, \bgamma}} = \frac{\cprob{\R_L}{\sigsq, \bgamma} \cprob{\R_R}{\sigsq, \bgamma}}{\cprob{\R}{\sigsq, \bgamma}} \\
%\eeqn
%
%Note that there's many things that can cancel above. The sum of the $R_i^2 / \sigsq_i$'s would cancel numerator-denominator since we're adding the same set\footnote{The set of left leaf values $\cup$ the set of right leaf values has to equal the set of the node's parent's values}, the product of the $\sigsq_i$'s would cancel since we're multiplying over the same set, the $2\pi$'s will cancel as well leaving us with:
%
%\begin{changemargin}{-20px}{0px}
%
\beqn
\frac{\cprob{\R}{\treet{*} \sigsq, \bgamma}}{\cprob{\R}{\treet{}, \sigsq, \bgamma}} &=& \sqrt{\frac{1 + \sigsq_\mu \displaystyle \sum_{i=1}^{n_\ell} \oneover{\sigsq_i}}{\parens{1 + \sigsq_\mu \displaystyle \sum_{i=1}^{n_{\ell_L}} \oneover{\sigsq_i}}\parens{1 + \sigsq_\mu \displaystyle \sum_{i=1}^{n_{\ell_R}} \oneover{\sigsq_i}}}} \\
&& \times~ \exp{\overtwo{\sigsq_\mu} \parens{\frac{\squared{\displaystyle \sum_{i=1}^{n_{\ell_L}} \frac{R_i}{\sigsq_i}}}{1 + \sigsq_\mu \displaystyle \sum_{i=1}^{n_{\ell_L}} \oneover{\sigsq_i}} + \frac{\squared{\displaystyle \sum_{i=1}^{n_{\ell_R}} \frac{R_i}{\sigsq_i}}}{1 + \sigsq_\mu \displaystyle \sum_{i_R=1}^{n_{\ell_R}} \oneover{\sigsq_i}} - \frac{\squared{\displaystyle \sum_{i=1}^{n_\ell} \frac{R_i}{\sigsq_i}}}{1 + \sigsq_\mu \displaystyle \sum_{i=1}^{n_\ell} \oneover{\sigsq_i}}}}.
\eeqn
%\end{changemargin}

%In log form this becomes:
%
%\inblue{
%\beqn
%&& \half \parens{\natlog{1 + \sigsq_\mu \displaystyle \sum_{i=1}^{n} \oneover{\sigsq_i}} - \natlog{1 + \sigsq_\mu \displaystyle \sum_{i_L=1}^{n_L} \oneover{\sigsq_i}} - \natlog{1 + \sigsq_\mu \displaystyle \sum_{i_R=1}^{n_R} \oneover{\sigsq_i}}} + \\
%&& \overtwo{\sigsq_\mu} \parens{\frac{\squared{\displaystyle \sum_{i_L=1}^{n_L} \frac{R_i}{\sigsq_i}}}{1 + \sigsq_\mu \displaystyle \sum_{i_L=1}^{n_L} \oneover{\sigsq_i}} + \frac{\squared{\displaystyle \sum_{i_R=1}^{n_R} \frac{R_i}{\sigsq_i}}}{1 + \sigsq_\mu \displaystyle \sum_{i_R=1}^{n_R} \oneover{\sigsq_i}} - \frac{\squared{\displaystyle \sum_{i=1}^{n} \frac{R_i}{\sigsq_i}}}{1 + \sigsq_\mu \displaystyle \sum_{i=1}^{n} \oneover{\sigsq_i}}}
%\eeqn
%}

\subsubsection*{The Likelihood Ratio for the \texttt{CHANGE} proposal}

The homoskedastic BART implementation of \citet{Kapelner2013} considered change proposals for singly internal nodes only (i.e., both daughter nodes must be terminal nodes). The likelihood ratio in this case simplifies to the likelihood of these two leaves before and after the change:

\newcommand{\Rlstar}[1]{R_{\ell^*, #1}}
\newcommand{\Rrstar}[1]{R_{r^*, #1}}
\newcommand{\Rl}[1]{R_{\ell, #1}}
\newcommand{\Rr}[1]{R_{r, #1}}
\newcommand{\nlstar}{n_{\ell^*}}
\newcommand{\nrstar}{n_{r^*}}

\bneqn\label{eq:hetero_change}
\frac{\cprob{\Rlstar{1}, \ldots, \Rlstar{\nlstar}}{\sigsq, \bgamma}\cprob{\Rrstar{1}, \ldots, \Rrstar{\nrstar}}{\sigsq, \bgamma}}{\cprob{\Rl{1}, \ldots, \Rl{n_\ell}}{\sigsq, \bgamma}\cprob{\Rr{1}, \ldots, \Rr{n_r}}{\sigsq, \bgamma}}.
\eneqn

\noindent The $\ell$ refers to the left terminal node and $r$ refers to the terminal right node. The $\ell^*$ and the $r^*$ denote these same two nodes in the proposal tree, i.e after the parent's split rule was changed.

The likelihood with $\mu$ margined out has been calculated in Equation~\ref{eq:marginalization} and we express it here with a convenient factorization:

\beqn
\underbrace{{\tothepow{(2\pi)^{n_\ell} \parens{1 + \sigsq_\mu \displaystyle \sum_{i=1}^{{n_\ell}} \oneover{\sigsq_i}} \displaystyle \prod_{i=1}^{n_\ell} \sigsq_i}{-\half}}}_{A} \underbrace{\exp{-\half \sum_{i=1}^{n_\ell}{\frac{R_i^2}{\sigsq_i}}}}_{B} \underbrace{\exp{\overtwo{\sigsq_\mu} \parens{\frac{\squared{\displaystyle \sum_{i=1}^{n_\ell} \frac{R_i}{\sigsq_i}}}{1 + \sigsq_\mu \displaystyle \sum_{i=1}^{n_\ell} \oneover{\sigsq_i}}}}}_{C}.\\
\eeqn

\noindent To find the ratio, we must substitute this expression into equation \ref{eq:hetero_change} four times. We begin by substituting only the term marked $A$ above to arrive at

\beqn
\frac{\tothepow{2\pi}{\overtwo{n_\ell}}\sqrt{\parens{\displaystyle \prod_{i=1}^{n_\ell} \sigsq_i} \parens{1 + \sigsq_\mu \displaystyle \sum_{i=1}^{n_\ell} \oneover{\sigsq_i}}} \tothepow{2\pi}{\overtwo{n_r}}\sqrt{\parens{\displaystyle \prod_{i=1}^{n_r} \sigsq_i} \parens{1 + \sigsq_\mu \displaystyle \sum_{i=1}^{n_r} \oneover{\sigsq_i}}}}{\tothepow{2\pi}{\overtwo{\nlstar}}\sqrt{\parens{\displaystyle \prod_{i=1}^{\nlstar} \sigsq_i} \parens{1 + \sigsq_\mu \displaystyle \sum_{i=1}^{\nlstar} \oneover{\sigsq_i}}} \tothepow{2\pi}{\overtwo{\nrstar}}\sqrt{\parens{\displaystyle \prod_{i=1}^{\nrstar} \sigsq_i} \parens{1 + \sigsq_\mu \displaystyle \sum_{i=1}^{\nrstar} \oneover{\sigsq_i}}}}.
\eeqn

\noindent Of course $\nlstar + \nrstar = n_\ell + n_r = n$ and the $\sigsq$'s are the same since they aren't drawn until later on in the Gibbs sampling scheme. Thus, the above reduces to

\bneqn\label{eq:term_A}
\sqrt{\frac{
\parens{
1 + \sigsq_\mu \displaystyle \sum_{i=1}^{n_\ell} \oneover{\sigsq_i}
}
\parens{
1 + \sigsq_\mu \displaystyle \sum_{i=1}^{n_r} \oneover{\sigsq_i}
}
}{
\parens{
1 + \sigsq_\mu \displaystyle \sum_{i=1}^{\nlstar} \oneover{\sigsq_i}
} 
\parens{
1 + \sigsq_\mu \displaystyle \sum_{i=1}^{\nrstar} \oneover{\sigsq_i}
}
}}.
\eneqn

\noindent In term $B$, upon making all four substitutions, it is clear all the parents' observations must be summed in the numerator as well as the denominator. Thus, this term cancels.

Now we examine term $C$. Due to the exponentiation, multiplication becomes addition and division becomes subtraction and all four substitutions yield

\bneqn\label{eq:term_C}
\exp{\overtwo{\sigsq_\mu} \parens{\frac{\squared{\displaystyle \sum_{i=1}^{\nlstar} \frac{R_i}{\sigsq_i}}}{1 + \sigsq_\mu \displaystyle \sum_{i=1}^{\nlstar} \oneover{\sigsq_i}} 
+ \frac{\squared{\displaystyle \sum_{i=1}^{\nrstar} \frac{R_i}{\sigsq_i}}}{1 + \sigsq_\mu \displaystyle \sum_{i=1}^{\nrstar} \oneover{\sigsq_i}} 
- \frac{\squared{\displaystyle \sum_{i=1}^{n_\ell} \frac{R_i}{\sigsq_i}}}{1 + \sigsq_\mu \displaystyle \sum_{i=1}^{n_\ell} \oneover{\sigsq_i}}
- \frac{\squared{\displaystyle \sum_{i=1}^{n_r} \frac{R_i}{\sigsq_i}}}{1 + \sigsq_\mu \displaystyle \sum_{i=1}^{n_r} \oneover{\sigsq_i}}}}.
\eneqn

\noindent Multiplying terms \ref{eq:term_A} and \ref{eq:term_C} yields the likelihood ratio for the \texttt{CHANGE} proposal.

\end{document}